\documentclass[12pt,preprint]{aastex}

\shorttitle{Dartmouth Stellar Evolution Database}
\shortauthors{Dotter et al.}

\newcommand{\Ms}{\mathrm{M_{\odot}}}
\newcommand{\Ls}{\mathrm{L_{\odot}}}
\newcommand{\Rs}{\mathrm{R_{\odot}}}
\newcommand{\Teff}{\mathrm{T_{eff}}}
\newcommand{\feh}{\mathrm{[Fe/H]}}
\newcommand{\afe}{\mathrm{[\alpha/Fe]}}
\newcommand{\Ks}{\mathrm{K_s}}

\begin{document}

\title{The Dartmouth Stellar Evolution Database}

\author{Aaron Dotter and Brian Chaboyer}
\affil{Department of Physics and Astronomy, Dartmouth College, 6127 Wilder Laboratory, Hanover, NH 03755}

\author{Darko Jevremovi\'c\altaffilmark{1}}
\affil{Astronomical Observatory, Volgina 7, 11160 Belgrade, Serbia}
\altaffiltext{1}{Homer L. Dodge Department of Physics and Astronomy, University of Oklahoma, 440 West Brooks, Room 100, Norman, OK 73019-2061}

\author{Veselin Kostov and E. Baron\altaffilmark{2}}
\affil{Homer L. Dodge Department of Physics and Astronomy, University of Oklahoma, 440 West Brooks, Room 100, Norman, OK 73019-2061}
\altaffiltext{2}{Computational Research Division, Lawrence Berkeley National Laboratory, MS 50F-1650, 1 Cyclotron Rd, Berkeley, CA 94720}

\author{Jason W. Ferguson}
\affil{Physics Department, Wichita State University, Wichita, KS 67260-0032}

\slugcomment{Accepted for publication in ApJS}

\begin{abstract}
The ever--expanding depth and quality of photometric and spectroscopic observations 
of stellar populations increase the need for theoretical models in regions of 
age--composition parameter space that are largely unexplored at present. Stellar
evolution models that employ the most advanced physics and cover a wide range of 
compositions are needed to extract the most information from current observations 
of both resolved and unresolved stellar populations. The Dartmouth Stellar Evolution
Database is a collection of stellar evolution tracks and isochrones that spans a 
range of $\feh$ from --2.5 to +0.5, $\afe$ from --0.2 to +0.8 (for $\feh \leq$0) or 
+0.2 (for $\feh >$0), and initial He mass fractions from Y=0.245 to 0.40. Stellar 
evolution tracks were computed for masses between 0.1 and 4 $\Ms$ allowing isochrones
to be generated for ages as young as 250 Myr. 
For the range in masses where the core He flash occurs, separate He-burning 
tracks were computed starting from the zero age horizontal branch. The tracks and 
isochrones have been transformed to the observational plane in a variety of 
photometric systems including standard UBV(RI)$_C$, Str\"omgren uvby, the Sloan 
Digital Sky Survey ugriz, the Two Micron All Sky Survey JH$\Ks$, and Hubble Space 
Telescope ACS-WFC and WFPC2. 
The Dartmouth Stellar Evolution Database is accessible through
 a website\footnote{http://stellar.dartmouth.edu/$\sim$models/}
 where all tracks, isochrones, and additional files can be downloaded.  
\end{abstract}

\keywords{globular clusters: general --- open clusters: general --- stars: evolution }

\section{Introduction}

Recent photometric surveys such as the ACS Survey of Galactic Globular Clusters 
\citep{sar} have substantially increased the quality and volume of data available 
for the Galactic globular cluster system.  Despite the fact that this survey focused 
on metal poor stellar populations it was found that there was a need for stellar 
evolution models with $\feh > $0 \citep{sieg}. Many other photometric surveys such 
as the Two Micron All Sky Survey\footnote{http://pegasus.phast.umass.edu/} (2MASS) 
and the Sloan Digital Sky Survey\footnote{http://www.sdss.org/} (SDSS) and extensions 
such as SEGUE\footnote{http://segue.uchicago.edu/} underscore the need for stellar 
evolution libraries that span a wide range of compositions and photometric systems.

\citet{dotter} introduced a set of stellar evolution tracks, isochrones, and computer
 programs to accompany the ACS Survey of Galactic Globular Clusters.  
The set includes tracks with masses between 0.1 and 1.8 $\Ms$, isochrones with ages 
ranging from 2 to 15 Gyr, [Fe/H] from --2.5 to 0 with a variety of options in $\afe$ 
and initial He content (Y$_{init}$). In this paper, that set of models is expanded to 
include: (i) tracks and isochrones with [Fe/H]= +0.15, +0.3, and +0.5; (ii) higher 
mass tracks and isochrones with ages as low as 250 Myr for all compositions; 
(iii) a wider selection of photometric systems; and (iv) a web interface to three of 
the computer programs so that visitors can easily generate specific models without downloading 
the entire database and running the programs locally.

Collectively, these tracks, isochrones, and supporting computer programs constitute 
the Dartmouth Stellar Evolution Database. 
The stellar evolution code used to generate these models is reviewed in $\S$2.  
Details of the compositions and model grids are outlined in $\S$3 and the 
color--$\Teff$ transformations in $\S$4.  In $\S$5, the isochrones are compared to 
those of comparable isochrone libraries and photometry of Galactic open clusters 
and one globular cluster.  The results are summarized and discussed in $\S$6.

\section{The Dartmouth Stellar Evolution Program}
The Dartmouth Stellar Evolution Program (DSEP) has been described previously by 
\citet{bjork} and \citet{dotter}.  The following is a brief description of the physics 
employed by DSEP.

Partially inhibited atomic diffusion and gravitational settling of both He (Y) and 
metals (Z) are included as described by \citet{cha}. Diffusion is calculated based 
on the formalism of \citet{thoul} but inhibited in the outer 0.01 $\Ms$ where the 
diffusion coefficients are set to zero in the outermost 0.005 $\Ms$ and ramped from 
zero to full diffusion in the inner 0.05 $\Ms$.  This method produces results that 
are consistent with recent observations of diffusion in NGC~6397 \citep{korn}.

DSEP uses high temperature opacities from OPAL \citep{opal} and low temperature 
opacities based on \citet{ferg}. The OPAL tables were obtained for each of the 
different $\afe$ mixtures from the OPAL web interface
\footnote{http://www-phys.llnl.gov/Research/OPAL/} and the low temperature tables were
calculated by one of us (Ferguson).  The low temperature opacity tables have been made 
available via the World Wide Web\footnote{http://webs.wichita.edu/physics/opacity/}.  
Conductive opacities are taken from \citet{hubb}.

The basic equation of state (EOS) in DSEP is a general ideal gas EOS with the 
Debye-H\"uckel correction \citep{chakim} and is used for tracks with M $\geq 0.8 
\Ms$.  For lower mass tracks (M$< 0.8 \Ms$), the FreeEOS equation of state 
\citep{irwin} is used in the EOS4 configuration.

\begin{deluxetable}{ccccccccccc}
\tabletypesize{\footnotesize}
\tablecolumns{11}
\tablewidth{0pc}
\tablecaption{Minimum Mass ($\Ms$) for Convective Core Overshoot\label{conv}}
\tablehead{\colhead{}&\multicolumn{6}{c}{Y=0.245+1.54Z}&\multicolumn{2}{c}{Y=0.33}&\multicolumn{2}{c}{Y=0.4}\\
\colhead{$\feh$}&\colhead{$\afe$=--0.2}&\colhead{0}&\colhead{0.2}&\colhead{0.4}&\colhead{0.6}&\colhead{0.8}&\colhead{0}&\colhead{0.4}&\colhead{0}&\colhead{0.4}}
\startdata
--2.5 &   1.6 &   1.6 &   1.6 &   1.5 &   1.3 &   1.3 &   1.5 &   1.5 &   1.5 &   1.5 \\ 
--2.0 &   1.4 &   1.4 &   1.4 &   1.4 &   1.3 &   1.3 &   1.5 &   1.5 &   1.5 &   1.5 \\
--1.5 &   1.4 &   1.4 &   1.4 &   1.4 &   1.3 &   1.3 &   1.4 &   1.4 &   1.4 &   1.4 \\
--1.0 &   1.4 &   1.4 &   1.4 &   1.4 &   1.2 &   1.2 &   1.4 &   1.4 &   1.4 &   1.4 \\
--0.5 &   1.2 &   1.2 &   1.2 &   1.2 &   1.2 &   1.0 &   1.2 &   1.2 &   1.2 &   1.2 \\
  0.0 &   1.2 &   1.2 &   1.1 &   1.1 &   1.0 &   0.9 &   1.2 &   1.1 &   1.2 &   1.1 \\
 +0.15&\nodata&   1.1 &\nodata&\nodata&\nodata&\nodata&\nodata&\nodata&\nodata&\nodata\\
 +0.3 &   1.1 &   1.1 &   1.1 &\nodata&\nodata&\nodata&\nodata&\nodata&\nodata&\nodata\\
 +0.5 &   1.1 &   1.1 &   1.0 &\nodata&\nodata&\nodata&\nodata&\nodata&\nodata&\nodata\\
\enddata
\end{deluxetable}

DSEP treats convection with the standard mixing length theory and a solar-calibrated 
mixing length, $\alpha_{ML}$=1.938. The models include convective core overshoot 
according to the scheme developed by \citet{dem}. The underlying idea is that the 
amount of overshoot grows with the size of the convective core.  The amount of core 
overshoot is parametrized as a multiple of the pressure scale height ($\lambda_{P}$) 
and is a function of stellar mass and composition.  The minimum stellar mass receives 
0.05 $\lambda_{P}$ of overshooting, models with masses greater than the minimum by 
0.1 $\Ms$ receive 0.1 $\lambda_{P}$ of overshooting, and models with masses greater 
than or equal to the minimum plus 0.2 $\Ms$ receive 0.2 $\lambda_{P}$ of overshooting.  
For details on the minimum mass as a function of composition see Table \ref{conv}. 
Convective envelope undershooting is not considered in these models.

The surface boundary condition (BC) was derived from a grid of \texttt{PHOENIX} model 
atmospheres \citep{phxa,phxb} covering $\Teff$=2,000 to 10,000 K and log g=--0.5 to 5.5.
  The solar abundance scale used 
in these calculations was that of \citet[hereafter GS98]{gs98}. \texttt{PHOENIX} model atmosphere grids were
computed for the full range $\feh$=--2.5 to +0.5 and $\afe$=--0.2 to +0.8 covered by the
stellar evolution calculations. A small number of model atmospheres with enhanced He were 
also computed to gauge the importance of enhanced He to the physical structure and the 
spectrum. The gas pressure (and thus the surface BC) experiences a small change with 
enhanced He and this is accounted for in the surface BC.
 
\citet{cas} atmospheres were used for $\Teff >$ 10,000 K; \citet{cas} models were computed 
only for $\afe$=0 and +0.4.  Stellar evolution models with $\afe \leq$ +0.2 made use of 
surface BC derived from $\afe$ = 0 model atmospheres for $\Teff >$ 10,000 K.
For stellar evolution models with $\afe \geq$ +0.4 the $\afe$ = +0.4 boundary conditions
were used for $\Teff >$ 10,000 K.  The transition between the \texttt{PHOENIX} and \citet{cas}
were always smooth regardless of the composition.  The smooth transition is due to the stellar 
evolution models being relatively insensitive to the surface BC at higher $\Teff$ coupled with
the reduced sensitivity of the model atmospheres to abundance variations in the transitional
$\Teff$.

The surface BC was fit at the point in the atmospheric structure where T=$\Teff$ for 
a given model.  The gas pressure was extracted at T=$\Teff$ and added to the radiation 
pressure to satisfy the surface BC.  \citet{vand07} found that, at 
least in the vicinity of 1 $\Ms$, the choice of model atmosphere boundary condition 
fitting point (either at T=$\Teff$ or at an optical depth of $\tau$=100) has little 
influence on the resulting evolution.

The \texttt{PHOENIX} model atmosphere code employs its own set of physics that is
not identical to the physics used in DSEP, though there is some overlap.  In
particular, the low temperature opacities of \citet{ferg} are calculated with 
\texttt{PHOENIX} and this implies a level of consistency between the two codes.
\texttt{PHOENIX} employs a standard mixing length theory of convection with
mixing length $\alpha_{MLT}$ = 2.0 (whereas DSEP uses $\alpha_{MLT}$ = 1.938). 
Perhaps the largest difference between \texttt{PHOENIX} and DSEP in terms of 
stellar structure is the EOS: \texttt{PHOENIX} uses an ideal EOS that includes the 
contributions of a large number of elements and molecules while FreeEOS (used by 
DSEP) includes several non-ideal effects but accounts for fewer elements and only
molecules involving combinations of H and He.

The nuclear reaction rates are taken from \citet{adel} for most reactions.  The 
exceptions are the $^{14}$N(p,$\gamma$)$^{15}$O reaction \citep{imbr}, the 
triple-$\alpha$ reaction from the NACRE compilation \citep{nacre}, and the 
$^{12}$C($\alpha$,$\gamma$)$^{16}$O reaction \citep{kunz}. Neutrino cooling rates were 
taken from \citet{haft}.

\section{Model Grids and Supporting Programs}

Stellar evolution models were computed using one of six different heavy element 
mixtures with the distinguishing characteristic of each mixture being the level of 
$\alpha$-element enhancement. Opacities were generated for each of $\afe$= --0.2, 0, 
0.2, 0.4, 0.6, and 0.8.  The heavy element mass fractions of these mixtures are given 
in Table \ref{mix}.  The mixtures were calculated starting with the solar mixture of 
GS98 by adding the value of $\afe$ to the log(N+12) abundance 
of each of the following $\alpha$-elements: O, Ne, Mg, Si, S, Ca, and Ti; the 
abundance of Ar was not altered from its GS98 value in any of the mixtures. 

Overall, models were computed for 67 different X,Z pairs; the initial compositions 
are listed in Table \ref{init}. For $\feh >$ 0 the maximum level of 
$\alpha$-enhancement was chosen to be $\afe$=+0.2 in order to keep the total Z value 
below 0.1 since the opacity tables do not extend above Z=0.1.  The model atmosphere 
grid included only provided scaled-solar ($\afe$=0) models at $\feh$=+0.15. The He 
content of the models begins with Y=0.245 at $\feh$ = --2.5, consistent with 
\citet{sper}, and a He enrichment relation of $\Delta$Y/$\Delta$Z=1.54.  For 
$\feh \leq$0, additional models were computed with Y=0.33 and 0.40 (see Table \ref{init}).

Stellar evolution tracks were computed for stellar masses ranging from 0.1 to 4 
$\Ms$ in increments of 0.05 $\Ms$ (between 0.1 and 1.8 $\Ms$), 0.1 $\Ms$ (between 
1.8 and 3 $\Ms$), and 0.2 $\Ms$ (between 3 and 4 $\Ms$).  The tracks cover the 
evolution of the models from the fully convective pre-main sequence (pre-MS) to 
either the core He flash for M $\la$ 2 $\Ms$ or either 100 Gyr or the top of the 
white dwarf cooling curve (whichever came first) for M $\la$ 0.5 $\Ms$. For models 
that experience the core He flash, separate He burning tracks were evolved from the 
Zero Age Horizontal Branch (ZAHB) using the surface compositions and He core masses 
listed in Tables \ref{zahb} and \ref{core}, respectively. For models that either 
transition smoothly to core He burning (M $\ga$ 2 $\Ms$) or were evolved from the 
ZAHB, the He--burning evolution was followed until the onset of thermal pulsations 
on the asymptotic giant branch (AGB).

Isochrones were generated from these tracks with ages from 250 Myr to 15 Gyr in 
increments of: 50 Myr between 250 Myr and 1 Gyr, 250 Myr between 1 and 5 Gyr, 
and 500 Myr between 5 and 15 Gyr.  The isochrones extend from the lower MS to the 
tip of the red giant branch (RGB) or, for ages where the tracks permit it (MS mass 
above $\sim$2 $\Ms$), at or near the onset of the thermal pulsations on the AGB
\footnote{Due to the mass spacing of the stellar evolution tracks and the interpolation
used in isochrone generation, the isochrones that include the AGB do not always terminate 
at the `tip' of the AGB.  In this case, it is recommended to examine the stellar evolution 
tracks if the brightness of the stars at the onset of the TP-AGB stage is of interest.
This applies only to iscohrones with ages below 1 Gyr.  Isochrones with ages above 1 Gyr
that include only the RGB provide reasonably accurate estimates of the RGB tip as a 
function of age.}. The low mass 
limit of the isochrones depends on the age: the isochrones only include masses for 
which the models have reached the MS.  The lowest mass stars considered in this paper 
(0.1 $\Ms$) may require $\sim$1 Gyr before reaching the MS and therefore will not be 
present in isochrones for younger ages.  For example, the $\feh$=0 isochrones have the 
following minimum masses ($\Ms$): 0.43 at 250 Myr, 0.32 at 500 Myr, 0.22 at 750 Myr, 
and 0.11 at 1 Gyr. Evolutionary tracks are also provided and these include the pre-MS 
evolution for all masses.

All tracks and isochrones were transformed to the color-magnitude diagram (CMD) using 
each of the color-$\Teff$ transformations described in $\S$4.  The track files consist 
of lines indicating age, log $\Teff$, log g, log L/$\Ls$, followed by the absolute 
magnitude in each of the filters for a given photometric system. A theoretical track 
file is also available, this file contains age, log $\Teff$, log g, log L/$\Ls$, log 
R/$\Rs$, core He mass fraction (Y$_{core}$), core Z mass fraction (Z$_{core}$), the 
surface Z/X ratio, H--burning luminosity ($\Ls$), He--burning luminosity 
($\Ls$), He core mass, and C/O core mass. The isochrones list stellar mass, 
log $\Teff$, log g, log L/$\Ls$, and absolute magnitude in each of the 
filters for a given photometric system.

The collection of Fortran 77 programs introduced by \citet{dotter} have been updated 
to deal with the expanded compositions and photometric systems presented in this 
paper but are otherwise largely unchanged. See the appendices of \citet{dotter} for 
a detailed description of these programs and their output.

\begin{deluxetable}{ccccccc}
\tablecolumns{7}
\tablewidth{0pc}
\tablecaption{Mass Fractions of Z\label{mix}}
\tablehead{\colhead{}&\multicolumn{6}{c}{$\afe$=}\\
\colhead{Element}&\colhead{--0.2}&\colhead{0}&\colhead{0.2}&\colhead{0.4}&\colhead{0.6}&\colhead{0.8}}
\startdata
 C & 2.296E-01 & 1.721E-01 & 1.233E-01 & 8.490E-02 & 5.685E-02 & 3.731E-02\\ 
 N & 6.726E-02 & 5.041E-02 & 3.611E-02 & 2.487E-02 & 1.665E-02 & 1.093E-02\\ 
 O & 3.940E-01 & 4.680E-01 & 5.314E-01 & 5.800E-01 & 6.155E-01 & 6.403E-01\\ 
Ne & 9.044E-02 & 1.050E-01 & 1.165E-01 & 1.271E-01 & 1.349E-01 & 1.403E-01\\ 
Na & 2.773E-03 & 2.078E-03 & 1.489E-03 & 1.025E-03 & 6.866E-04 & 4.506E-04\\ 
Mg & 3.366E-02 & 3.998E-02 & 4.540E-02 & 4.955E-02 & 5.258E-02 & 5.470E-02\\ 
Al & 4.814E-03 & 3.608E-03 & 2.585E-03 & 1.780E-03 & 1.192E-03 & 7.822E-04\\ 
Si & 3.715E-02 & 4.412E-02 & 5.010E-02 & 5.468E-02 & 5.803E-02 & 6.036E-02\\ 
 P & 6.493E-04 & 4.866E-04 & 3.486E-04 & 2.401E-04 & 1.607E-04 & 1.055E-04\\ 
 S & 1.851E-02 & 2.199E-02 & 2.497E-02 & 2.725E-02 & 2.892E-02 & 3.008E-02\\ 
Cl & 3.900E-04 & 2.923E-04 & 2.094E-04 & 1.442E-04 & 9.656E-05 & 6.337E-05\\ 
Ar & 5.793E-03 & 4.342E-03 & 3.111E-03 & 2.142E-03 & 1.434E-03 & 9.414E-04\\ 
 K & 3.045E-04 & 2.282E-04 & 1.635E-04 & 1.126E-04 & 7.539E-05 & 4.948E-05\\ 
Ca & 3.268E-03 & 3.882E-03 & 4.408E-03 & 4.811E-03 & 5.106E-03 & 5.311E-03\\ 
Ti & 1.519E-04 & 1.804E-04 & 2.048E-04 & 2.236E-04 & 2.372E-04 & 2.468E-04\\ 
Cr & 1.470E-03 & 1.102E-03 & 7.894E-04 & 5.436E-04 & 3.640E-04 & 2.389E-04\\ 
Mn & 1.075E-03 & 8.054E-04 & 5.771E-04 & 3.974E-04 & 2.661E-04 & 1.746E-04\\ 
Fe & 1.020E-01 & 7.641E-02 & 5.474E-02 & 3.770E-02 & 2.524E-02 & 1.657E-02\\ 
Ni & 6.026E-03 & 4.516E-03 & 3.235E-03 & 2.228E-03 & 1.492E-03 & 9.791E-04
\enddata
\end{deluxetable}

\begin{deluxetable}{cllllllllll}
\rotate
\tabletypesize{\footnotesize}
\tablecolumns{11}
\tablewidth{0pc}
\tablecaption{Initial Compositions\label{init}}
\tablehead{\colhead{}&\multicolumn{6}{c}{Y=0.245+1.54Z}&\multicolumn{2}{c}{Y=0.33}&\multicolumn{2}{c}{Y=0.4}\\
\colhead{$\feh$}&\colhead{$\afe$=--0.2}&\colhead{0}&\colhead{0.2}&\colhead{0.4}&\colhead{0.6}&\colhead{0.8}&\colhead{0}&\colhead{0.4}&\colhead{0}&\colhead{0.4}}
\startdata
--2.5 & X=0.7549 & 0.7548 & 0.7548 & 0.7547 & 0.7546 & 0.7543 & 0.6699 & 0.6699 & 0.6000 & 0.5999 \\
\phn & Z=4.09E-5 & 5.48E-5 & 7.63E-5 & 1.11E-4 & 1.65E-4 & 2.52E-4 & 4.85E-5 & 9.85E-5 & 4.34E-5 & 8.82E-5 \\
--2.0 & X=0.7547 & 0.7545 & 0.7544 & 0.7541 & 0.7536 & 0.7529 & 0.6698 & 0.6697 & 0.5999 & 0.5997 \\
\phn & Z=1.30E-4 & 1.72E-4 & 2.41E-4 & 3.50E-4 & 5.24E-4 & 7.97E-4 & 1.53E-4 & 3.50E-4 & 1.37E-4 & 2.79E-4 \\
--1.5 & X=0.7539 & 0.7536 & 0.7530 & 0.7521 & 0.7507 & 0.7484 & 0.6695 & 0.6690 & 0.5996 & 0.5991 \\
\phn & Z=4.10E-4 & 5.47E-4 & 7.62E-4 & 1.11E-3 & 1.65E-3 & 2.50E-3 & 4.85E-4 & 9.83E-4 & 4.34E-4 & 8.81E-4 \\
--1.0 & X=0.7516 & 0.7505 & 0.7487 & 0.7459 & 0.7415 & 0.7346 & 0.6685 & 0.6669 & 0.5986 & 0.5972 \\
\phn & Z=1.29E-3 & 1.72E-3 & 2.40E-3 & 3.46E-3 & 5.15E-3 & 7.77E-3 & 1.53E-3 & 3.46E-3 & 1.37E-3 & 2.78E-3 \\
--0.5 & X=0.7444 & 0.7441 & 0.7355 & 0.7270 & 0.7139 & 0.6942 & 0.6652 & 0.6603 & 0.5957 & 0.5913 \\
\phn & Z=4.05E-3 & 5.37E-3 & 7.44E-3 & 1.07E-2 & 1.57E-2 & 2.32E-2 & 4.82E-3 & 9.70E-3 & 4.31E-3 & 8.69E-3 \\
 0.0 & X=0.7174 & 0.7071 & 0.6882 & 0.6628 & 0.6247 & 0.5710 & 0.6550 & 0.6402 & 0.5866 & 0.5734 \\
\phn & Z=1.43E-2 & 1.89E-2 & 2.55E-2 & 3.52E-2 & 4.97E-2 & 7.02E-2 & 1.50E-2 & 2.98E-2 & 1.34E-2 & 2.66E-2 \\
+0.15& X=\nodata & 0.6880 & \nodata&\nodata&\nodata&\nodata&\nodata&\nodata &\nodata&\nodata\\
\phn & Z=\nodata & 2.56E-2 & \nodata&\nodata&\nodata&\nodata&\nodata&\nodata &\nodata&\nodata\\
+0.3 & X=0.6842 & 0.6638 & 0.6334 & \nodata&\nodata&\nodata&\nodata&\nodata &\nodata&\nodata\\
\phn & Z=2.70E-2 & 3.48E-2 & 4.64E-2 & \nodata&\nodata&\nodata&\nodata&\nodata &\nodata&\nodata\\
+0.5 & X=0.6491 & 0.6185 & 0.5792 & \nodata&\nodata&\nodata&\nodata&\nodata &\nodata&\nodata\\
\phn & Z=4.04E-2 & 5.21E-2 & 6.71E-2 & \nodata&\nodata&\nodata&\nodata&\nodata &\nodata&\nodata\\
\enddata
\end{deluxetable}

\begin{deluxetable}{cllllllllll}
\rotate
\tabletypesize{\footnotesize}
\tablecolumns{11}
\tablewidth{0pc}
\tablecaption{Surface Compositions for ZAHB Models\label{zahb}}
\tablehead{\colhead{}&\multicolumn{6}{c}{Y$_{init}$=0.245+1.54Z$_{init}$\tablenotemark{\dagger}}&\multicolumn{2}{c}{Y$_{init}$=0.33}&\multicolumn{2}{c}{Y$_{init}$=0.4}\\
\colhead{$\feh$}&\colhead{$\afe$=--0.2}&\colhead{0}&\colhead{0.2}&\colhead{0.4}&\colhead{0.6}&\colhead{0.8}&\colhead{0}&\colhead{0.4}&\colhead{0}&\colhead{0.4}}
\startdata
--2.5 & X=0.7564  & 0.7565  & 0.7561  & 0.7549  & 0.7549  & 0.7546  & 0.6835  & 0.6828  & 0.5967  & 0.6212  \\
\phn & Z=3.97E-5 & 5.31E-5 & 7.40E-4 & 1.60E-4 & 1.08E-4 & 2.44E-4 & 4.63E-5 & 9.40E-5 & 8.40E-3 & 8.33E-5 \\

--2.0 & X=0.7547  & 0.7548  & 0.7551  & 0.7541  & 0.7524  & 0.7510  & 0.6820  & 0.6810  & 0.6208  & 0.6193  \\
\phn & Z=1.26E-4 & 1.67E-4 & 2.33E-4 & 3.41E-4 & 5.06E-4 & 7.69E-4 & 1.46E-4 & 2.97E-4 & 1.30E-4 & 2.64E-4 \\

--1.5 & X=0.7529  & 0.7517  & 0.7509  & 0.7491  & 0.7464  & 0.7427  & 0.6785  & 0.6763  & 0.6180  & 0.6156  \\
\phn & Z=3.96E-4 & 5.29E-4 & 7.37E-4 & 1.07E-3 & 1.60E-3 & 2.43E-3 & 4.64E-4 & 9.44E-4 & 4.12E-4 & 8.37E-4 \\

--1.0 & X=0.7472  & 0.7453  & 0.7423  & 0.7383  & 0.7325  & 0.7238  & 0.6732  & 0.6698  & 0.6125  & 0.6076  \\
\phn & Z=1.25E-3 & 1.67E-3 & 2.33E-3 & 3.37E-3 & 5.02E-3 & 7.59E-3 & 1.47E-3 & 3.00E-3 & 1.31E-3 & 2.67E-3 \\  

--0.5 & X=0.7350  & 0.7309  & 0.7242  & 0.71485 & 0.7010  & 0.6810  & 0.6641  & 0.6600  & 0.6034  & 0.5967  \\
\phn & Z=3.94E-3 & 5.24E-3 & 7.27E-3 & 1.05E-2 & 1.54E-3 & 2.28E-2 & 4.67E-3 & 9.48E-3 & 4.16E-3 & 8.41E-3 \\  
 
 0.0 & X=0.7166  & 0.6928  & 0.6741  & 0.6497  & 0.6131  & 0.5640  & 0.6477  & 0.6303  & 0.5877  & 0.5854  \\
\phn & Z=1.41E-2 & 1.85E-2 & 2.50E-2 & 3.46E-2 & 4.88E-2 & 6.87E-2 & 1.47E-2 & 2.91E-2 & 1.31E-2 & 2.68E-2 \\

+0.15& X=\nodata  & 0.6737  &\nodata&\nodata&\nodata&\nodata&\nodata&\nodata &\nodata&\nodata\\
\phn & Z=\nodata  & 2.51E-2 &\nodata&\nodata&\nodata&\nodata&\nodata&\nodata &\nodata&\nodata\\

+0.3 & X=0.6696  & 0.6505  & 0.6204  &\nodata&\nodata&\nodata&\nodata&\nodata &\nodata&\nodata\\
\phn & Z=2.66E-2 & 3.42E-2 & 4.57E-2 &\nodata&\nodata&\nodata&\nodata&\nodata &\nodata&\nodata\\

+0.5 & X=0.6360  & 0.6075  & 0.5734  & \nodata&\nodata&\nodata&\nodata&\nodata &\nodata&\nodata\\
\phn & Z=3.98E-2 & 5.12E-2 & 6.57E-2 & \nodata&\nodata&\nodata&\nodata&\nodata &\nodata&\nodata\\
\enddata
\tablenotetext{\dagger}{Y$_{init}$ and Z$_{init}$ refer to the initial Y and Z values, for the initial 
compositions see Table \ref{init}.}
\end{deluxetable}

\begin{deluxetable}{ccccccccccc}
\tabletypesize{\footnotesize}
\tablecolumns{11}
\tablewidth{0pc}
\tablecaption{He Core Mass ($\Ms$) for ZAHB Models\label{core}}
\tablehead{\colhead{}&\multicolumn{6}{c}{Y=0.245+1.54Z}&\multicolumn{2}{c}{Y=0.33}&\multicolumn{2}{c}{Y=0.4}\\
\colhead{$\feh$}&\colhead{$\afe$=--0.2}&\colhead{0}&\colhead{0.2}&\colhead{0.4}&\colhead{0.6}&\colhead{0.8}&\colhead{0}&\colhead{0.4}&\colhead{0}&\colhead{0.4}}
\startdata
--2.5 & 0.506 & 0.505 & 0.504 & 0.502 & 0.500 & 0.497 & 0.489 & 0.485 & 0.474 & 0.473 \\
--2.0 & 0.500 & 0.499 & 0.498 & 0.494 & 0.492 & 0.491 & 0.483 & 0.479 & 0.469 & 0.468 \\
--1.5 & 0.495 & 0.493 & 0.491 & 0.489 & 0.488 & 0.486 & 0.478 & 0.474 & 0.464 & 0.463 \\
--1.0 & 0.489 & 0.488 & 0.486 & 0.484 & 0.481 & 0.478 & 0.473 & 0.469 & 0.460 & 0.458 \\
--0.5 & 0.483 & 0.482 & 0.480 & 0.477 & 0.473 & 0.469 & 0.469 & 0.465 & 0.457 & 0.455 \\
  0.0 & 0.475 & 0.473 & 0.469 & 0.464 & 0.458 & 0.447 & 0.465 & 0.460 & 0.454 & 0.451 \\
 +0.15&\nodata& 0.468 &\nodata&\nodata&\nodata&\nodata&\nodata&\nodata&\nodata&\nodata\\
 +0.3 & 0.468 & 0.464 & 0.459 &\nodata&\nodata&\nodata&\nodata&\nodata&\nodata&\nodata\\
 +0.5 & 0.461 & 0.456 & 0.449 &\nodata&\nodata&\nodata&\nodata&\nodata&\nodata&\nodata\\
\enddata
\end{deluxetable}

\section{Color-$\Teff$ Transformations}
Synthetic color-$\Teff$ transformations were derived from the \texttt{PHOENIX} model 
atmosphere grid (also used to generate surface boundary conditions in the stellar 
evolution calculations) for $\Teff <$ 10,000 K and for all $\feh$ and $\afe$
combinations. The \texttt{PHOENIX} grid was supplemented by \citet{cas} models for higher 
$\Teff$.  In order to insure a smooth transition, the \citet{cas} colors were adjusted 
to line up with the \texttt{PHOENIX} colors at 10,000 K and the two sets of tables were 
ramped between \texttt{PHOENIX} and \citet{cas} in the temperature range of 9,000 K to 10,000 K.

As mentioned in $\S$2, the \texttt{PHOENIX} model atmospheres used to generate 
synthetic color--$\Teff$ transformations were computed for each of the $\feh$, $\afe$
pairs but not for all of the enhanced Y values. The influence of enhanced He on 
broad-band synthetic photometry is negligible for the temperatures considered here, 
as discussed briefly by \citet{dotter} and in more detail by \citet{gir07}.

The \citet{cas} model atmospheres were only computed for $\afe$=0 and +0.4.  In
constructing the color transformations the same approach was used as described in 
$\S$2 for the surface BC.  For evolutionary tracks and isochrones with 
$\afe \leq +0.2$, colors based on \citet{cas} synthetic spectra with $\afe = 0$ were applied and 
for tracks and isochrones with $\afe \geq +0.4$, colors based on \citet{cas} synthetic spectra with
$\afe = +0.4$ were applied.  While this is not an ideal situation, the influence of 
$\alpha$-enhancement, or indeed of total Z, on broad-band synthetic colors is small
for $\Teff \ga$ 5,000 K.  The only exceptions are in the bluest bands (equivalent to U)
and at $\Teff \ga$ 35,000 K.  Bands equivalent to U are the most uncertain in terms 
of the synthetic photometry and very few of the evolutionary tracks (only the lowest
mass, metal poor He-burning tracks) presented in this paper extend beyond 
$\Teff$ = 35,000 K.

The synthetic colors have been tested and perform well in bandpasses equivalent to 
V or redder (with central wavelengths longer than $\sim5000$ \AA) but the blue and 
ultraviolet bands (with central wavelengths shorter than $\sim5000$ \AA) suffer 
from inaccuracy of the synthetic fluxes at shorter wavelengths.  In cases where 
analysis in the bluer bands is important, empirical color transformations are
strongly recommended.

\citet{cass} explored the dependence of broadband synthetic colors on $\afe$ for
$\afe$=0 and +0.4. Their findings indicate that U--B and B--V are
the colors most strongly effected by $\alpha$-enhancement and that the influence of 
$\alpha$-enhancement on broadband colors increases with increasing metallicity
and decreasing $\Teff$. The color transformations derived from the 
\texttt{PHOENIX} model atmosphere grid have a greater span of $\Teff$ and $\afe$, 
including $\afe$ = --0.2 for the first time, and better resolution (0.2 dex) in $\afe$.
The comparisons performed with the \texttt{PHOENIX} synthetic colors qualitatively
agree with the findings of \citet{cass} though it is important to recognize that
\citet{cass} compared colors at constant Z while in this paper comparisons are
made at constant $\feh$.  Hence an increase in $\afe$ represents an increase in 
total Z and thus one expects the colors to become redder with increasing $\afe$
as a general rule.

\begin{figure}
\plotone{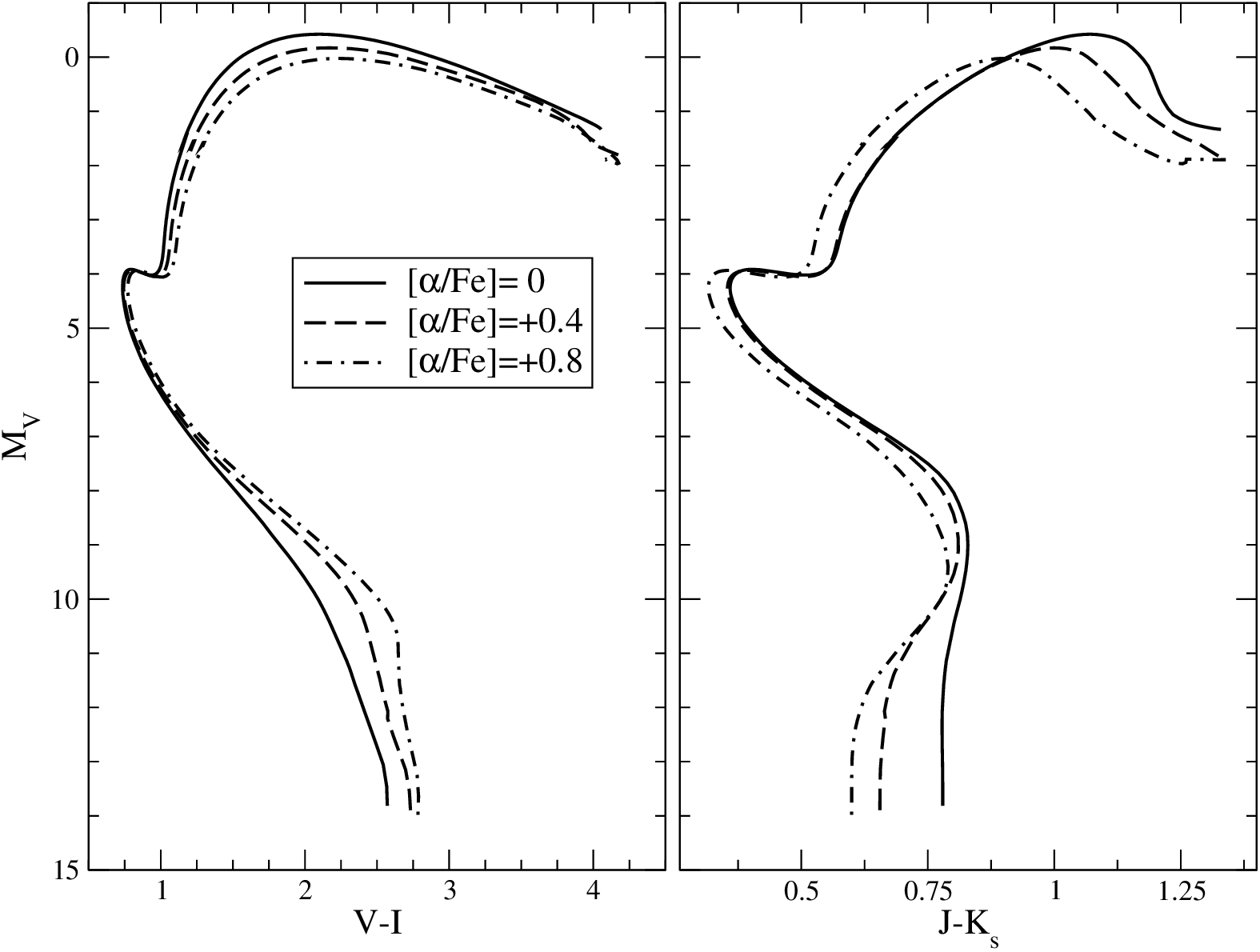}
\caption{The influence of $\alpha$-enhancement on broadband V--I and J--$\Ks$ colors at $\feh$ = 0.  The underlying isochrones shown are all scaled-solar composition with $\feh$ = 0 at 10 Gyr. Only the color transformations include $\alpha$-enhancement.  See the text for a discussion of the trends shown in the figure.}
\label{alpha}
\end{figure}

Figure \ref{alpha} displays the changes to CMD morphology based solely on the level
of $\alpha$-enhancement at $\feh$=0 using a 10 Gyr scaled-solar isochrone.  The
figure shows $\afe$ = 0, +0.4, and +0.8 colors.  The comparisons were assembled using
a scaled-solar isochrone in order to demonstrate the effect of
$\alpha$-enhancement on the colors only.  Two colors, V--I and J--$\Ks$ are shown
in Figure \ref{alpha}. The figure shows that V--I increases with $\afe$ and 
while the difference is small near the MSTO, it grows as $\Teff$ decreases, 
becoming largest near M$_V$ = 10.  The situation in J--$\Ks$ is reversed: the 
highest level of $\afe$ produces the bluest J--$\Ks$ color while the lowest
level produces the reddest color.  The J--$\Ks$ result is, at first glance,
counter-intuitive.  However, the J band is particularly sensitive to 
absorption from water and, in the presence of enhanced oxygen (from $\afe$), substantially 
more water is able to form.  This pushes flux away from the J band and causes the
J--$\Ks$ color to become bluer with increasing $\afe$.

As reported by \citet{cass}, the color differences due to variations in $\afe$ 
decrease as total metallicity decreases. However, isochrones presented in this 
paper suggest that the separation along the lower MS in J--$\Ks$ remains noticeable 
even at $\feh$=--2 and this feature may prove a useful diagnostic for near infrared 
studies of metal poor systems of sufficient depth to see the lower MS.

The remainder of this section give a brief description of the photometric systems
into which the stellar evolution tracks and isochrones have been transformed.

\subsection{Johnson--Cousins UBVRI and the Two Micron All Sky Survey}
\citet{bes} defined filter transmission curves for the Johnson--Cousins UBV(RI)$_C$ 
photometric system. These filter curves were combined with 2MASS JH$\Ks$ filters from 
\citet{cohen} to create a set of isochrones suitable for use with ground--based 
photometry from the ultraviolet to the near infrared. The UBVRI magnitudes were 
calibrated to Vega by following Appendix A of \citet{bcp}.  The 2MASS magnitudes were 
calibrated by setting the magnitudes of Vega to zero in each filter \citep{cohen}.

\subsection{The Hubble Space Telescope}
\citet{bed} defined a method for generating synthetic photometry based on the 
detailed filter transmission curves for the Hubble Space Telescope (HST) Advanced 
Camera for Surveys (ACS) and Wide Field Planetary Camera 2 (WFPC2) from \citet{sir}. 
The synthetic photometry presented here are all on the Vegamag system.  The following 
filters are included for the ACS-WFC: F435W, F475W, F555W, F606W, F625W, F775W, F814W, 
and F850LP. The following filters are included for the WFPC2: F336W, F439W, F555W, 
F606W, F791W, F814W, and F850LP. The performance of the synthetic color transformation 
in the ACS F606W-F814W CMD has been documented by \citet{sar,dotter,rich,sieg}.

\subsection{The Sloan Digital Sky Survey}
SDSS produced (and its ongoing extensions continue to produce) vast quantities of 
stellar photometry and spectroscopy with important implications for our knowledge of 
galactic structure and evolution. The SDSS ugriz photometric system is based on the 
AB magnitude system but each of the five filters has an additional offset from AB of 
at most a few hundredths of a magnitude. The offsets, in the sense of 
mag(SDSS)--mag(AB) are: 0.04 in u, 0.01 in g, 0 in r, --0.015 in i, and --0.03 in z. 
The filter transmission curves and procedure for transforming synthetic fluxes to SDSS 
magnitudes magnitudes is outlined on the SDSS Photometric Flux Calibration web page
\footnote{http://www.sdss.org/dr6/algorithms/fluxcal.html}.

Typically, the synthetic colors provide good fits to SDSS data in g, r, and i.
Synthetic photometry in the u band is subject to the same shortcomings as other bands
at bluer wavelengths while problems with the underlying calibration of the SDSS 
photometric systems persist in both u and z.

\subsection{Semi--Empirical Transformations}
For general use, and comparison with the synthetic color transformations, the tracks and
isochrones are presented in colors based on the semi--empirical transformations of 
\citet[hereafter VC03]{vdb} in BV(RI)$_C$.  VC03 compiled a set
of $\Teff$--color transformations and bolometric corrections covering a wide range of 
metallicity, $\Teff$, and log g.  The VC03 colors are based on MARCS model atmospheres 
\citep{bg78,bg89,vb85} for lower temperatures and Kurucz models (consistent with \citet{cas99})
for higher temperatures.  
The color transformations are semi--empirical in the sense that VC03 began with a 
collection of synthetic colors and then applied corrections to the colors, primarily
for cooler MS stars approaching and above $\feh$=0, based on observational constraints
from open clusters, globular clusters, and field stars.

\citet{clem} presented color transformations for the Str\"omgren uvby photometric system.
\citet{clem} used a similar approach to VC03 by adopting MARCS models \citep{houd} for 
$\Teff < $ 8,000 K and colors based on Kurucz models \citep{cgk97} for higher temperatures. 
The uvby color transformations of \citet{clem} provide much better fits to observations 
than any synthetic color tables (see that paper for a detailed discussion) and thus no 
synthetic Str\"omgren photometry has been provided.

\section{Comparisons}
This section will compare the DSEP models to comparable models and observational data 
from the literature.  \citet{dotter} presented comparisons with old, metal--poor 
isochrones (12 Gyr, Z=0.0001) from several groups and compared the DSEP isochrones to 
ACS photometry of two globular clusters: M~92 and 47~Tuc from the ACS Galactic 
Globular Cluster Survey \citep{sar}. As a result, this section will focus on model 
comparisons at $\feh \geq 0$, along with younger, more metal rich open clusters.
The one exception is a comparison of SDSS photometry from M~13 and NGC~2420 to the 
ugriz isochrones in $\S$5.2.6.

\subsection{Comparisons with Other Models}
Following \citet{dotter}, comparison isochrones were drawn from \citet[hereafter 
BaSTI]{piet,piet2,cord}, \citet[hereafter Padova]{gir00}, \citet[hereafter 
Victoria-Regina (VR)]{vdb2}, and \citet[hereafter Yale-Yonsei (Y$^2$)]{yi2,yi}. 
In this section, the comparison is extended to solar metallicity and above as well as 
to younger ages.  Comparisons are shown in both the log $\Teff$-log L/$\Ls$ plane and 
the V vs. V--I color-magnitude diagram.  In each case, the DSEP isochrone is shown 
with the synthetic color transformation.  A comparison of different color 
transformations applied to the DSEP isochrones appears later in this section.

\begin{figure}
\plotone{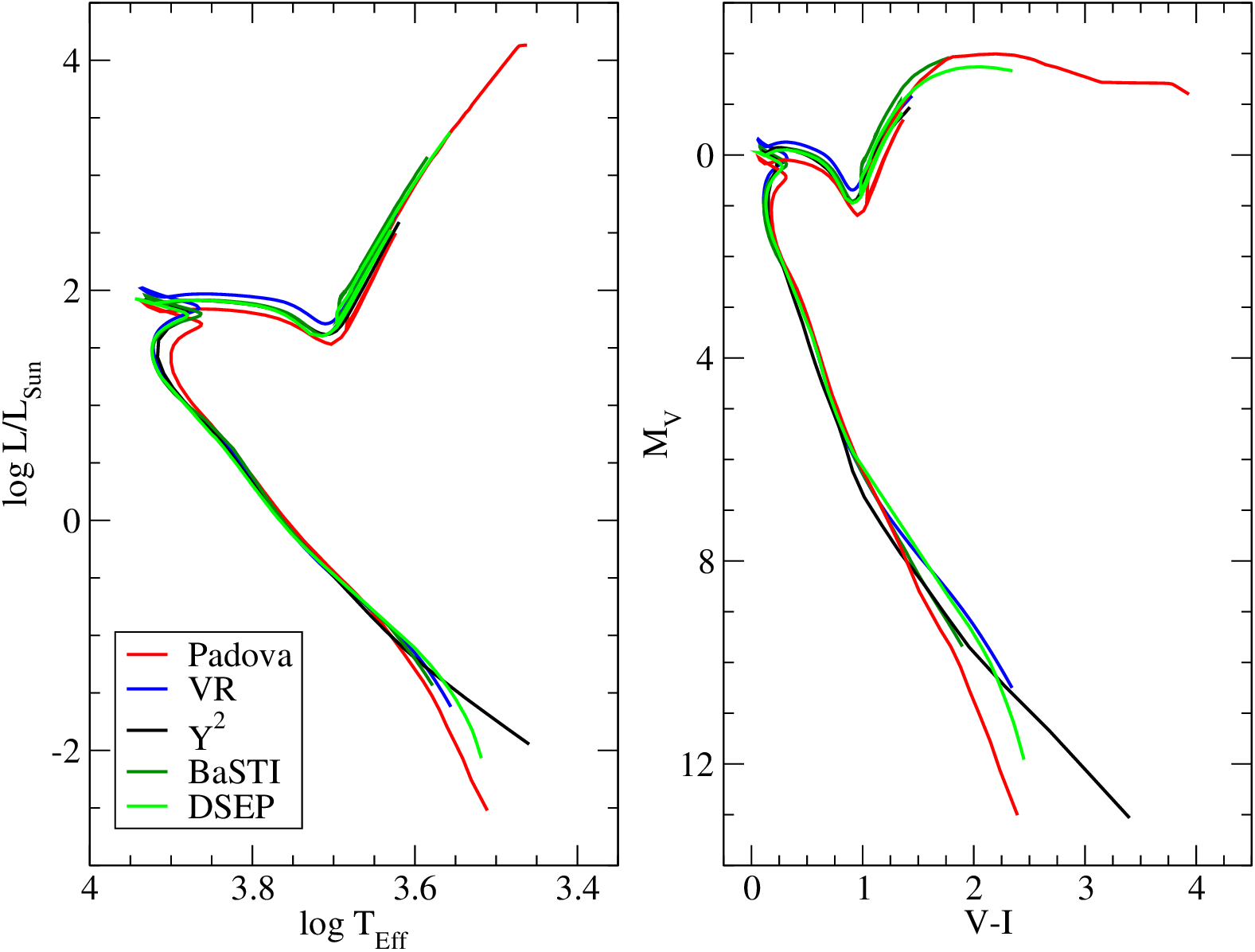}
\caption{A plot of isochrones at $\feh \sim$+0.15 and an age of $\sim$ 625 Myr.  
The Padova, BaSTI, and DSEP isochrones show the He burning phase but the others do 
not. Good agreement exists among all the isochrones except for Padova that is cooler 
and fainter around the main sequence turn off.}
\label{600Myr}
\end{figure}

\begin{figure}
\plotone{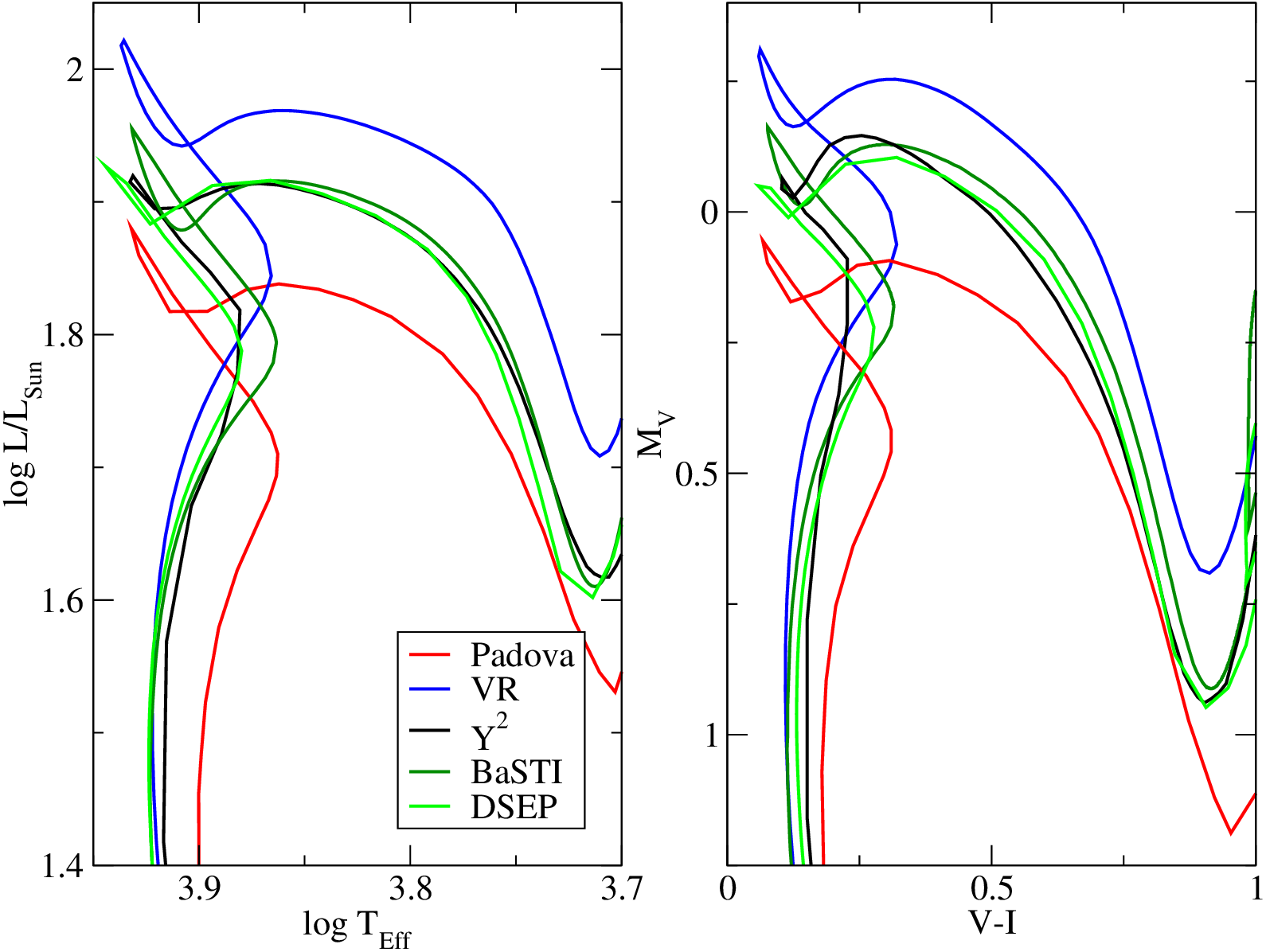}
\caption{The same as Figure \ref{600Myr} but focused on the region near the main sequence turn off.}
\label{600MSTO}
\end{figure}

Figures \ref{600Myr} and \ref{600MSTO} show isochrones at $\feh \sim$+0.15 and an age of 625-630 Myr 
(compatible with the the Hyades, see $\S$5.2.2). The isochrones range in initial 
Z=0.025 to 0.03 and Y=0.28 to 0.3.  In Figure \ref{600Myr}, the Padova, BaSTI, 
and DSEP isochrones all show the AGB though the point at which the AGB terminates 
differs: DSEP and BaSTI stop at or before the onset of thermal pulsations though 
the BaSTI library has been updated to include extended AGB evolution \citep{cord}.
Fainter than log L/$\Ls$=--1, the large differences are due to different mass ranges, 
the EOS, and low temperature opacities used by each group. The BaSTI isochrone begins 
at 0.5 $\Ms$, VR and Y$^2$ isochrones begin at 0.4 $\Ms$, the Padova isochrone begins 
at 0.15 $\Ms$, and the DSEP isochrone begins at 0.3 $\Ms$ at this age. 

In Figure \ref{600MSTO} the agreement is quite good among DSEP, BaSTI, and Y$^2$ isochrones. 
The Padova isochrone is less luminous in the region of the MS and subgiant 
branch (SGB) while the VR isochrons is more luminous in the same region.  These differences
in luminosity are primarily due to different adopted treatment of convective core overshoot.

Figure \ref{600Myr} indicates that the Y$^2$ isochrone extends to cooler 
temperatures than any of the others despite the fact that it terminates at 0.4 $\Ms$ 
while DSEP and Padova terminate at considerably lower mass. This effect is almost 
certainly due to the EOS used in the Y$^2$ calculations.

\begin{figure}
\plotone{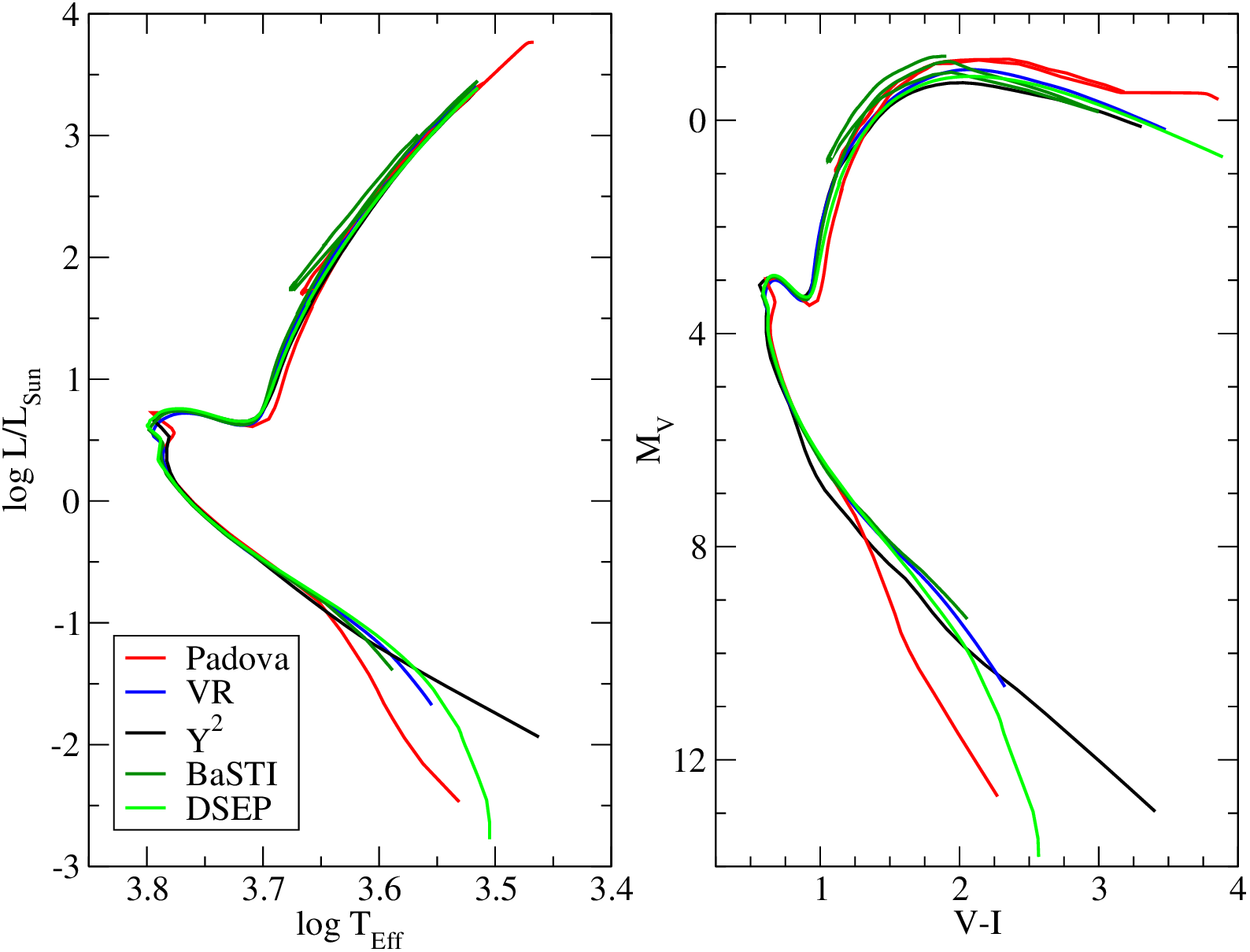}
\caption{The figure shows solar metallicity isochrones at 4 Gyr.  The Padova and 
BaSTI isochrones include the He burning phase but the others do not.  There is 
remarkable agreement in the H--R diagram between the tip of the red giant branch and 
log L/$\Ls$=--1. Below this point, the range of masses plotted and the adopted EOS 
cause large differences.  There is good agreement in the color-magnitude diagram 
between M$_V$=6 and 0 but beyond this the mass range plotted, the underlying physics, 
and the choice of color transformation cause significant discrepancies.}
\label{4gyr}
\end{figure}

\begin{figure}
\plotone{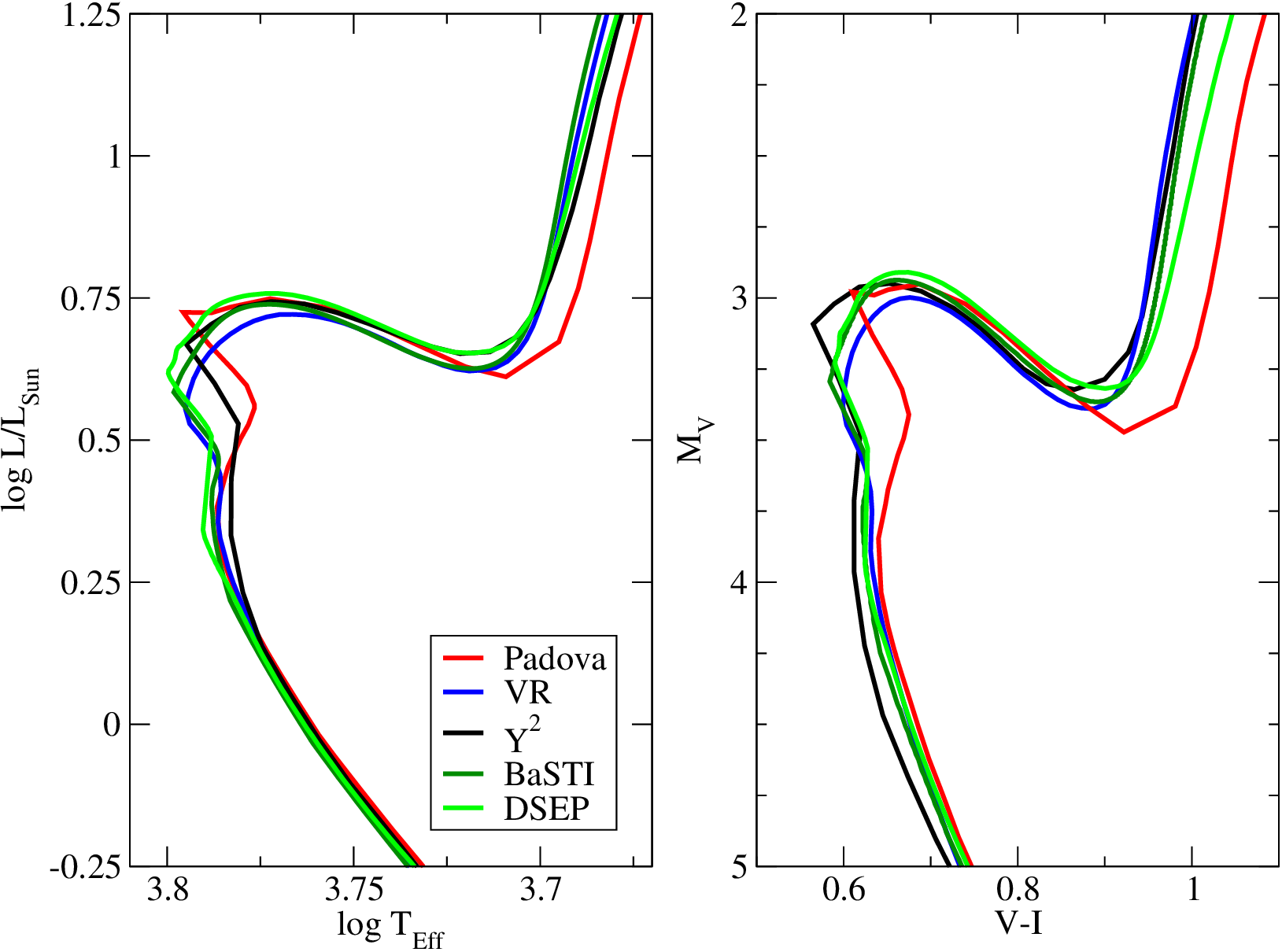}
\caption{The same as Figure \ref{4gyr} but focused on the region near the main sequence turn off.}
\label{4MSTO}
\end{figure}

Figures \ref{4gyr} and \ref{4MSTO} show isochrones at solar metallicity and 4 Gyr (compatible with 
M~67, see $\S$5.2.4).  Each isochrone has an initial Z=0.018-0.020 and Y=0.27-0.28; 
the adopted physics (especially the EOS and convective core overshoot prescription) 
and solar heavy element mixture differ from one group to another. In the left panels 
of Figure \ref{4gyr} and \ref{4MSTO}, the isochrones are in reasonable agreement for points brighter 
than log L/$\Ls$=--1 (Padova and BaSTI isochrones include He burning stars while the 
others do not).  In Figure \ref{4MSTO}, the BaSTI, VR, and DSEP isochrones all agree well 
on the temperature and luminosity of the MS turn off (TO) and SGB. In the right panels
of Figure \ref{4gyr} and \ref{4MSTO}, all isochrones agree well at magnitudes brighter than M$_V$=6 
except for Padova that is consistently redder and brighter on the upper RGB. In Figure
 \ref{4gyr} the DSEP isochrone includes the complete lower mass range (down to 0.1 
$\Ms$) and the fact that the Y$^2$ isochrone is cooler and almost as faint in M$_V$ is 
significant. Conversely, the Padova isochrone is hotter and bluer than the others for 
points fainter that log L/$\Ls$=--1 (M$_V$=7) although the mass-luminosity relation is 
similar to the DSEP isochrone.

Figures \ref{600Myr}, \ref{600MSTO}, \ref{4gyr} and \ref{4MSTO} show general agreement 
amongst the different sets
of isochrones.  The main exceptions are the Padova isochrones that appear cooler and 
redder near the MSTO and on the RGB but hotter and bluer on the lower main sequence.
Amongst the other models the largest differences occur near the MSTO where the adopted
core overshoot treatments differ and on the lower main sequence where the adopted EOS
and minimum masses differ.  The DSEP models use the same EOS and opacities as the VR
and BaSTI models and it is reassuring that these models all lie close together aside 
from the noted exceptions due to the included mass range and treatment of convective
core overshoot.

\subsection{Comparisons with Observational Data}
Several papers have demonstrated the performance of the DSEP isochrones when applied 
to HST/ACS data in two of the most popular bands for globular cluster photometry 
(F606W and F814W) including: \citet{sar}, \citet{dotter}, \citet{sieg}, and 
\citet{rich}.  This section will focus on the ground-based photometric systems 
presented in this paper and also compares the synthetic color transformation with the 
VC03 transformation in B--V and V--I.

\begin{deluxetable}{ccccccc}
\tablecolumns{7}
\tablewidth{0pc}
\tablecaption{Comparison Clusters\label{clusters}}
\tablehead{\colhead{ID}&\colhead{DM$_V$}&\colhead{E(B--V)}&\colhead{$\feh$}&\colhead{Approximate Age (Gyr)}&\colhead{Photometry}&\colhead{Reference(s)}}
\startdata
M~37     & 11.55 & 0.27  & 0.09 & 0.5 & B V J $\Ks$ & 1,2 \\
Hyades   &\nodata&\nodata& 0.13 & 0.65& B V I $\Ks$ & 3,4 \\
NGC~2420 & 12.1  & 0.05  &--0.27 & 2   & B V I $\Ks$ & 1   \\
M~67     &  9.65 & 0.04  & 0    & 4   & B V I $\Ks$ & 1,5,6 \\
NGC~6791 & 13.35 & 0.1   & 0.4  & 9   & B V I $\Ks$ & 7,8,9 \\
\enddata
\tablerefs{1--GS03, 2--\citet{kalirai}, 3--\citet{deb}, 4--\citet{pinson}, 5--\citet{mmj}, 6--{taylor}, 7--\citet{carney}, 8--\citet{stetson},9--\citet{cha3}}
\end{deluxetable}

\citet[hereafter GS03]{groch} combined optical photometry of six Galactic open 
clusters with near-infrared data from 2MASS. Similar compilations for the Hyades 
\citep{pinson} and NGC~6791 \citep{carney} are presented along with three clusters 
from GS03 to span a range of ages and metallicities in open clusters.  These combined 
optical and near-infrared data provide an excellent source of comparisons with the 
DSEP isochrones. 
Table \ref{clusters} lists the relevant parameters of the clusters featured in this 
section.

The adopted reddening values are listed as E(B--V). The extinction curve of 
\citet{johnson} was used to determine the reddening in other filter combinations:
E(V--I) = 1.4 E(B--V), E(V--J) = 2.4 E(B--V) and E(V-$\Ks$) = 2.8 E(B--V).  

These comparisons are not meant to determine the ages of the clusters but rather to 
demonstrate the performance of the isochrones. The ages were chosen to be consistent 
with other published results but if the age shown is considerably different from the 
true age of the cluster it will influence the quality of the fit.

\subsubsection{M~37}
\begin{figure}
\plotone{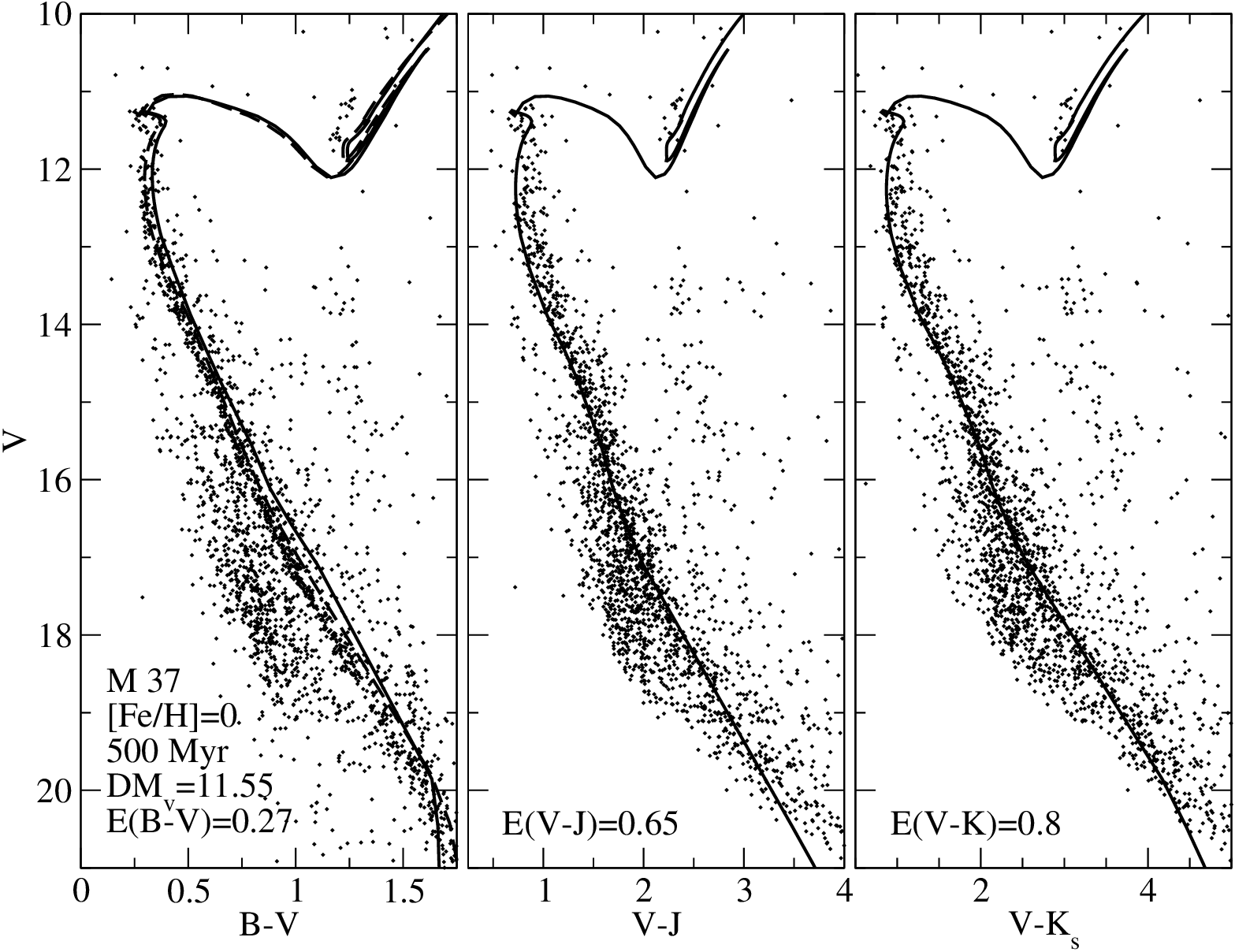}
\caption{CMDs of M~37 from \citet{kalirai} and GS03 are plotted with a $\feh$=0, 
500 Myr isochrone. The isochrone is shown in synthetic (solid line) and VC03 
(dashed line) colors. The synthetic color isochrone is redder than the data in 
B--V but a good fit in V--J and V--$\Ks$.  The He burning portion
of the isochrones is redder than the data in all cases.}
\label{m37}
\end{figure}

Figure \ref{m37} shows the CMD of M~37 in B--V \citep{kalirai} and V--J and V--$\Ks$ 
(GS03) along with a $\feh$=0, 500 Myr isochrone.  The isochrone is plotted using the 
synthetic (solid line) and VC03 (dashed line) colors.

In B--V (left panel of Figure \ref{m37}), the synthetic color isochrone is redder 
than the data by $\sim$0.1 mag throughout the CMD while the VC03 isochrone provides 
a reasonable fit to the data from the MSTO to V=16 before drifting to the red of the 
data.  The He burning portion of the isochrone is redder than the data by 0.1 mag in 
both color transformations.

In V--J and V--$\Ks$, the synthetic color isochrone provides an excellent fit to the 
data for the entire extent of the MS (the MS data cuts off at V$\sim$20 in V--J and 
V--$\Ks$).  The isochrone is redder than the data for the He burning stars in both 
V--J and V--$\Ks$.

We note in passing that the VC03 B--V isochrone shown in the left panel of figure
\ref{m37} does not match the shape of the CMD for V $>$ 17.  Although M~37 and M~67
are both roughly $\feh$ = 0 clusters their B--V CMDs from \citet{kalirai} and M~67 \citep{mmj}, 
respectively, adjusted for distance and reddening, 
appear to disagree somewhat for M$_V \ga$ 6 when viewed together.  Since the VC03 colors are 
based in part on the CMD of M~67, it is impossible for them to fit both clusters simultaneously.

\subsubsection{The Hyades}

\begin{figure}
\plotone{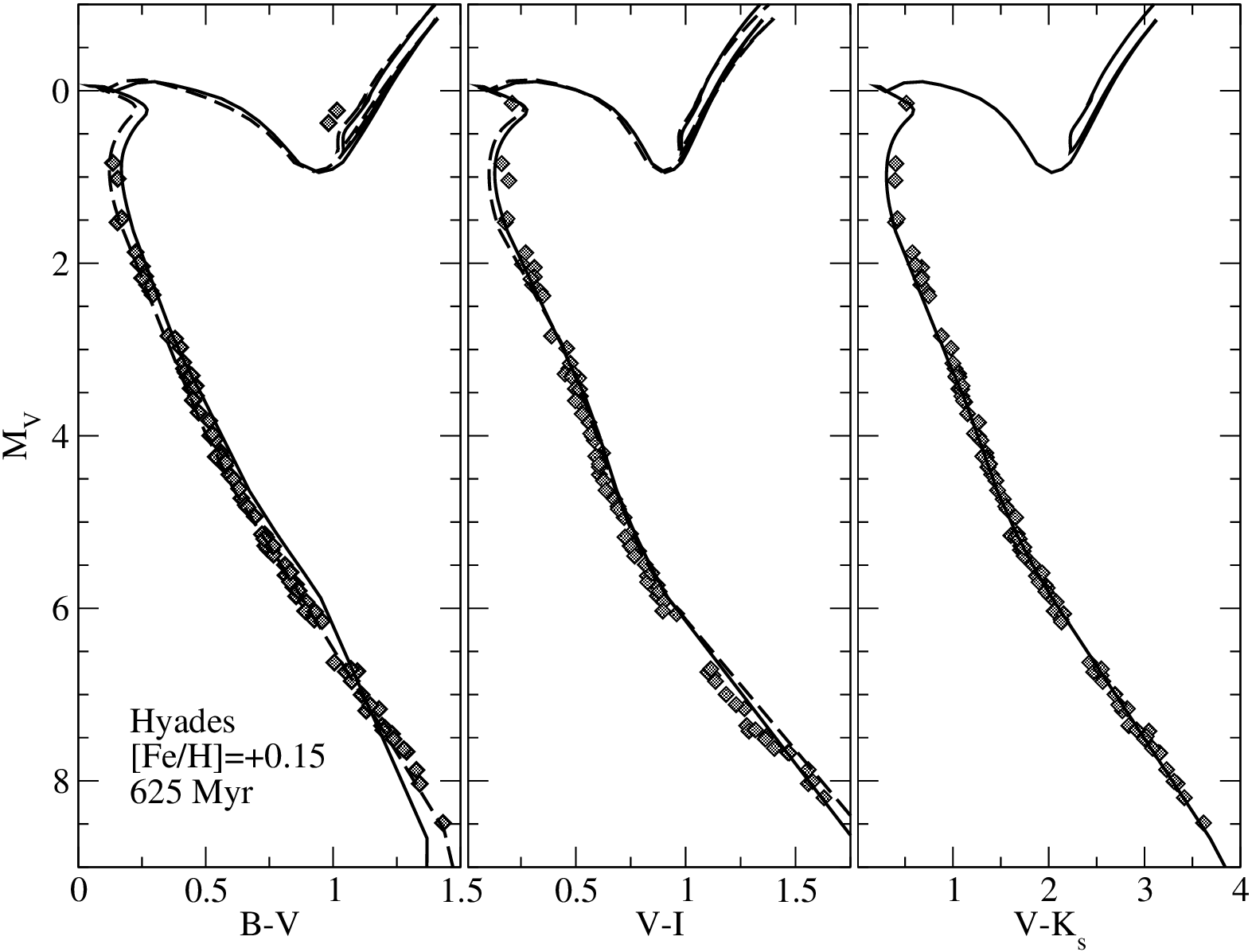}
\caption{The Hyades in B--V and V--I from \citet{deb} along with V--$\Ks$ from 
\citet{pinson}. Along with the data, an isochrone with $\feh$=+0.15 at 625 Myr is 
plotted using the synthetic colors (solid line) and VC03 colors (dashed line).}
\label{hya}
\end{figure}

Figure \ref{hya} shows the Hipparcos B--V (left panel) and V--I (center) CMDs of 
the Hyades \citep{deb} and the combined Hipparcos and 2MASS V--$\Ks$ CMD compiled 
by \citet{pinson}. Along with the data, a $\feh$=+0.15, 625 Myr isochrone is shown 
in the synthetic (solid line) and VC03 (dashed line) color transformations.  The 
choice of a 625 Myr isochrone may be an underestimate of the age which is typically
quoted as between 650 and 700 Myr (e.g. VC03). The He abundance assumed for the 
$\feh$=+0.15 models may also be a cause for disagreement since the adopted 
$\Delta$Y/$\Delta$Z yields Y$_{init}$=0.2864 whereas the Hyades Y value inferred from 
model comparisons is lower by $\sim$0.02 (see VC03 and references therein for a 
discussion).

In B--V, the synthetic color isochrone is redder than the data by 0.05--0.1 mag down 
to M$_V$=7 where the slope changes and the isochrone becomes bluer than the data by 
M$_V$=8. The VC03 color isochrone, on the other hand, is a good fit for the entirety 
of the MS.  Both isochrones appear cooler than the two He burning stars in the B--V 
CMD.

In V--I, the color transformations are quite similar and both perform well down to 
M$_V$=6. At this point, the isochrones become redder than the data before matching 
the data again at M$_V$=8.

In V--$\Ks$, the synthetic color isochrone fits the data for stars with M$_V \geq$ 3 
but becomes bluer than the data just below the MSTO. The discrepancy just below the 
MSTO would be avoided by plotting a slightly older isochrone.
 
\subsubsection{NGC~2420}

\begin{figure}
\plotone{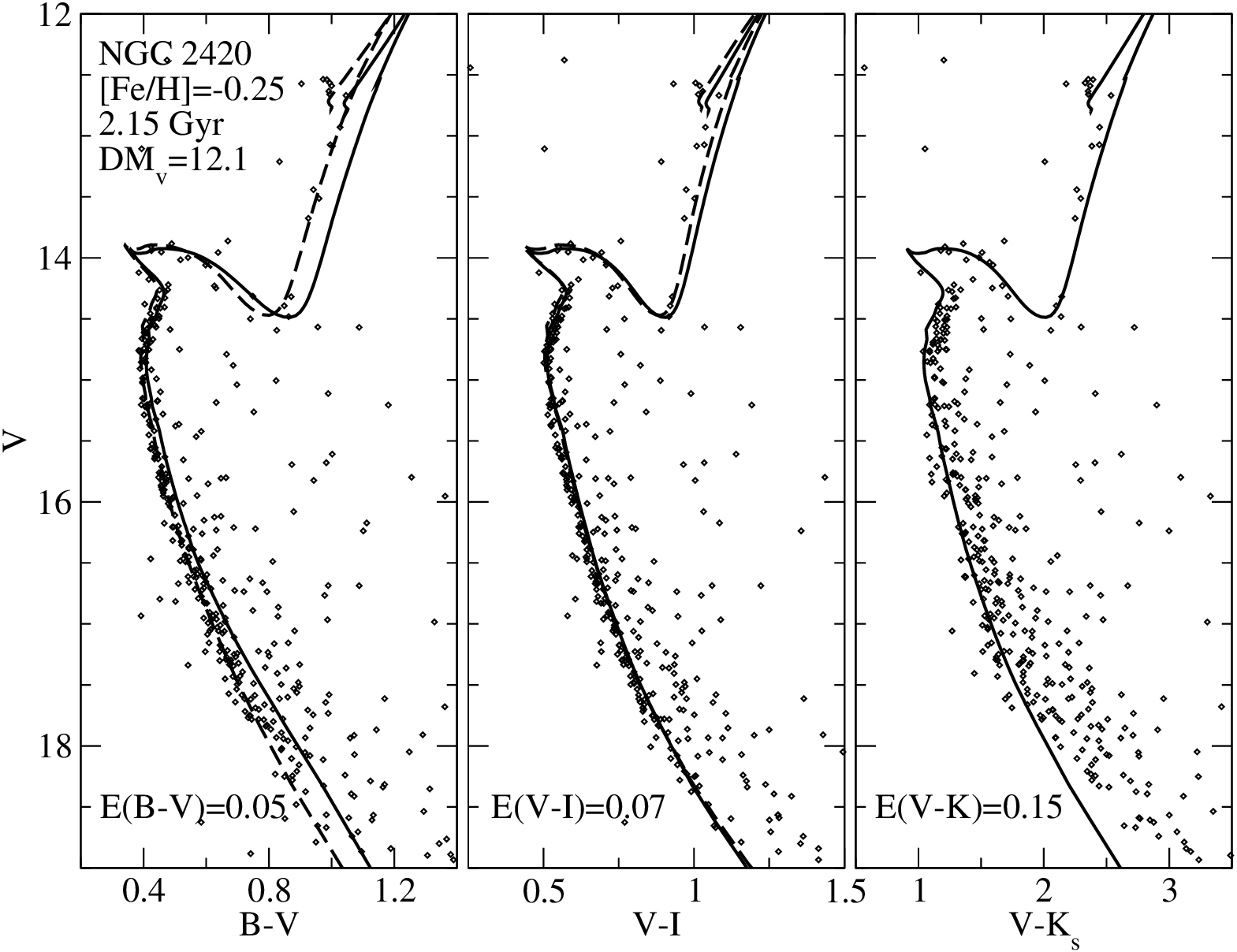}
\caption{NGC~2420 from GS03 along with an isochrone interpolated to $\feh$=--0.25 
and 2.15 Gyr. The isochrone is shown with synthetic colors (solid line) and VC03 
colors (dashed line).  Also shown is a 1.5 $\Ms$ He burning track interpolated to 
$\feh$=--0.25. }
\label{n2420}
\end{figure}

Figure \ref{n2420} shows the B--V, V--I, and V--$\Ks$ CMDs of NGC~2420 from GS03.  
Plotted along with the data is an isochrone interpolated to $\feh$=--0.25 and 2.15 
Gyr.  Given the adopted distance and reddening in Table \ref{clusters}, the best
fit age plotted in Figure \ref{n2420} is older than the age quoted by GS03.  This
is not surprising since ultimately the GS03 age estimate for NGC~2420 was based
on Padova isochrones and these were shown to be fainter and redder near the MSTO 
than the DSEP isochrones (see Figures \ref{600Myr} and \ref{4gyr} and the 
discussion in $\S$5.1).  As a result, the best fit Padova isochrone will be
younger than the best fit DSEP isochrone.

In B--V the synthetic isochrone is redder than the data, particularly on the lower 
MS and the RGB, though it matches well near the MSTO. The VC03 color isochrone fits 
the entire CMD well. The He burning track lies fainter than the red clump for both 
color transformations with the color predicted correctly by VC03 but too red by the 
synthetic color. If NGC~2420 is younger than 2.1 Gyr, then the stars in the red clump 
will be more massive and more luminous.

In V--I, both synthetic and VC03 colors fit the CMD well except for the synthetic 
color on the RGB that is slightly cooler than the data. The color of the red clump 
is correctly predicted by VC03 but slightly too red in the synthetic color.

In V--$\Ks$, the synthetic isochrone fits the SGB and RGB and correctly predicts 
the color of the red clump.  However, the V--$\Ks$ color appears to be too blue on 
the MS.

\subsubsection{M~67}

\begin{figure}
\plotone{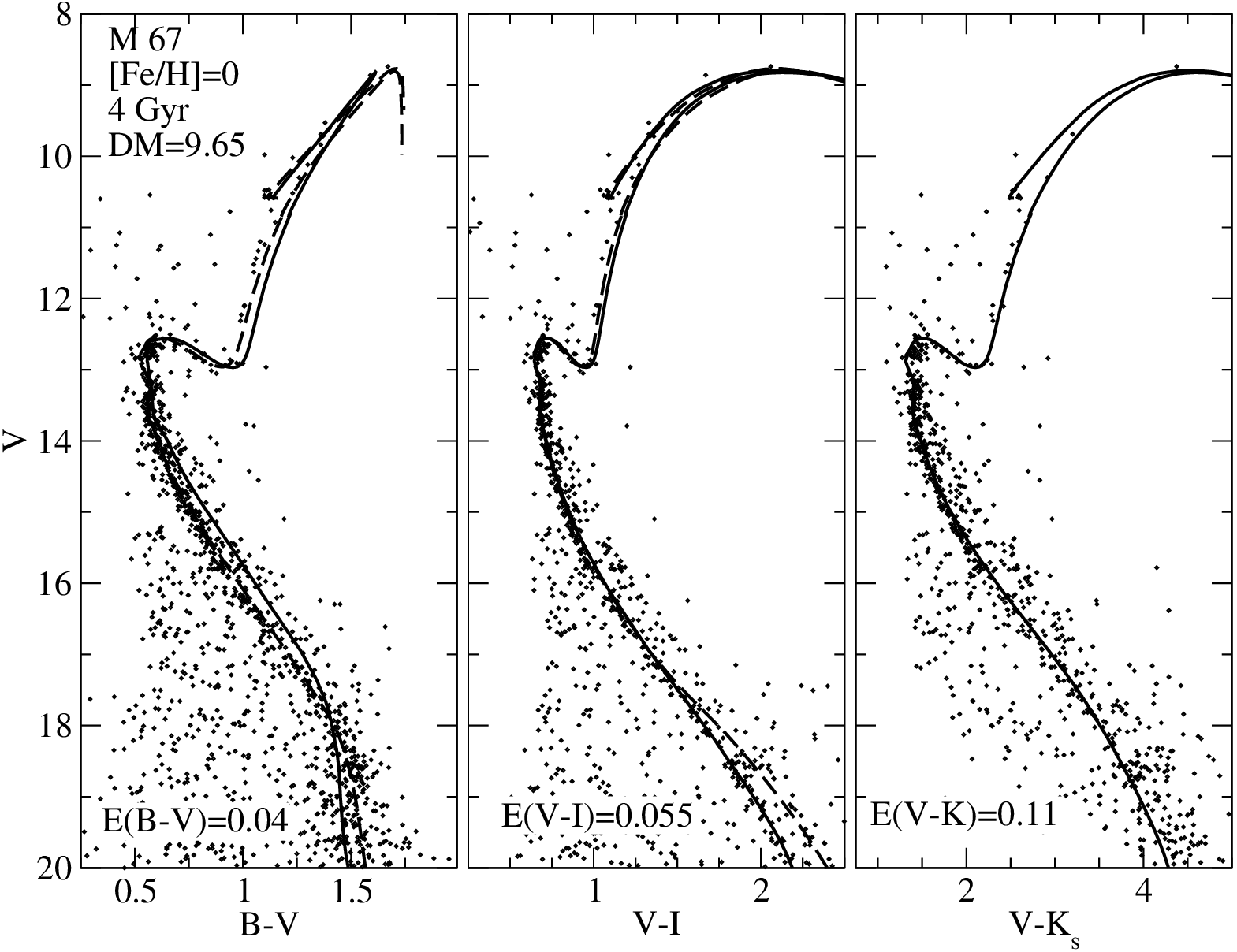}
\caption{The CMDs of M~67 in B--V and V--I \citep{mmj} and V--$\Ks$ (GS03).  
Isochrones with $\feh$=0 at 4 Gyr are shown using synthetic (solid line) and 
VC03 (dashed line) colors. A He-burning track with M=1.25 $\Ms$ at $\feh$=0 is 
plotted to indicate the predicted level of the red clump.}
\label{m67}
\end{figure}

Figure \ref{m67} shows the B--V, V--I, and V--$\Ks$ CMDs of M~67 from \citet{mmj} 
and GS03 along with a $\feh$=0, 4 Gyr isochrone. The isochrone has been transformed 
using the synthetic (solid line) and VC03 (dashed line) colors. A He burning track 
with $\feh$=0 and M=1.25 $\Ms$ is shown to indicate the predicted location of the 
red clump.

The fit of the synthetic color isochrone to the data in B--V (left panel of Figure 
\ref{m67}) is poor, with the isochrone redder than the data by about 0.05-0.1 mag 
throughout the CMD while the VC03 color isochrone provides an excellent fit. The 
He burning model has the correct V magnitude but is slightly 
redder than the observed red clump in B--V for both color transformations.

In V--I (middle panel of Figure \ref{m67}), both synthetic and VC03 colors match 
the data well over most of the CMD.  The only exception is that, for stars fainter 
than V=17, the VC03 colors are redder than the data while the synthetic colors 
follow the data closely down to V=20.  The He burning model correctly predicts the
color of the red clump in V--I for both color transformations.

In V--$\Ks$ only the synthetic color isochrone is plotted in the right panel of 
Figure \ref{m67}.  As with V--I, the synthetic V--$\Ks$ color is an excellent fit 
to the data for the lower MS, the MSTO, the RGB, and the location of the red clump.

\subsubsection{NGC~6791}

\begin{figure}
\plotone{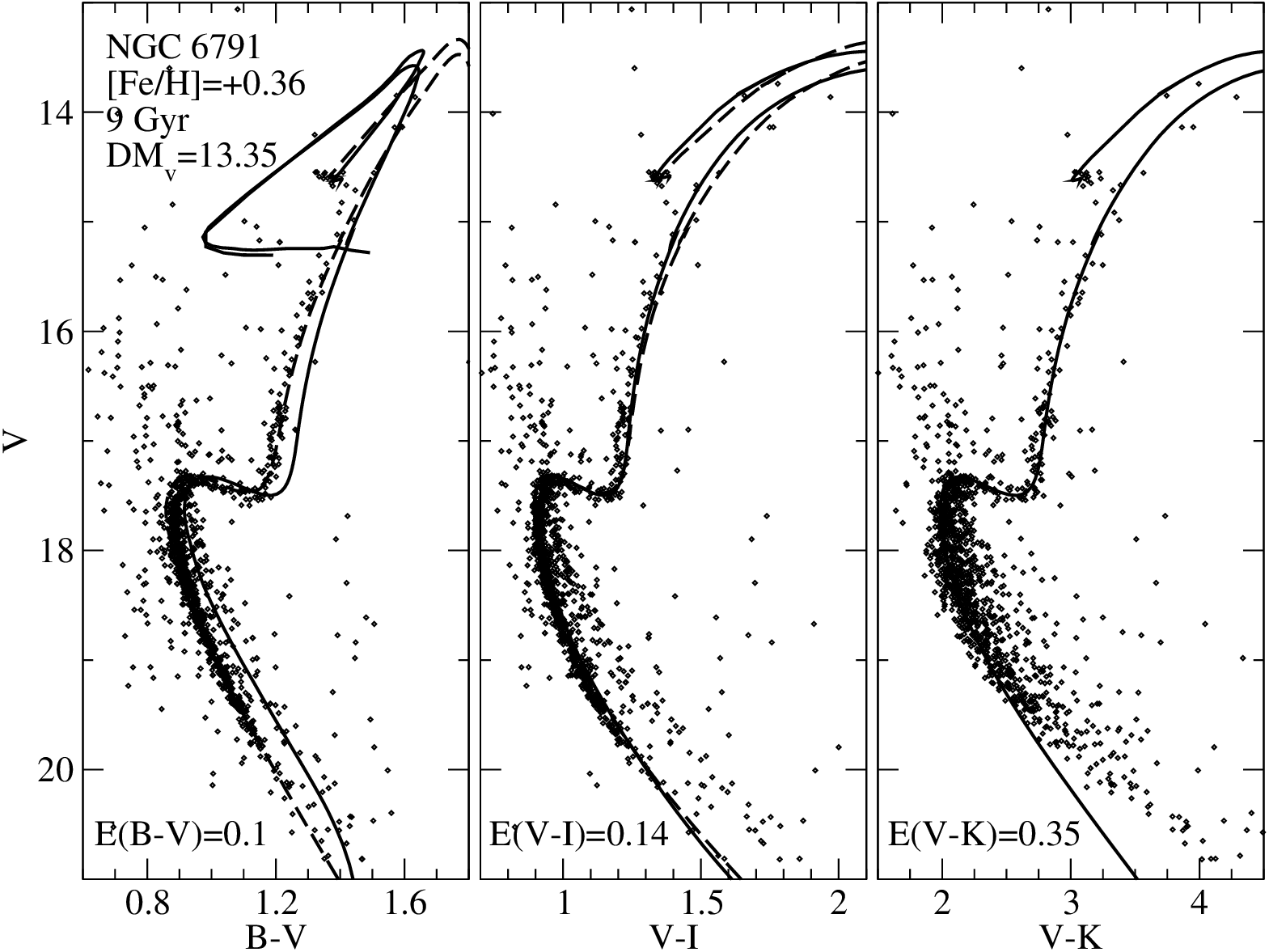}
\caption{The B--V, V--I, V--K CMDs of NGC~6791 from \citet{stetson} and 
\citet{carney}. Plotted on top of the data are isochrones with $\feh$=+0.36 at 
9 Gyr in synthetic (solid line) and VC03 (dashed line) colors. A 1 $\Ms$ He burning 
track of the same composition is plotted to indicate the predicted location of the 
red clump.}
\label{n6791}
\end{figure}

\citet{carney} combined JHK photometry of NGC~6791 with optical data from 
\citet{stetson}. The resulting CMDs in B--V, V--I, and V--K are shown in Figure 
\ref{n6791}. The V--K CMD has been adjusted by +0.024 mag to convert from CIT K to 
2MASS $\Ks$ \citep{carpenter}.  The isochrones shown in Figure \ref{n6791} have 
$\feh_{init}$ = +0.36 and an age of 9 Gyr.

The B--V CMD continues to be problematic for the synthetic colors where, using the 
adopted E(B--V), the synthetic isochrone is $\sim$0.1 mag redder than the data.  The 
isochrone with VC03 colors is a substantially better fit to the data.  This is not 
surprising since the VC03 colors were calibrated at high metallicity with the 
\citet{stetson} data.  The V--I CMD is equally well fit by both synthetic and VC03 
colors.  The synthetic color isochrone in V--K is an excellent fit to the shape of 
the CMD as well.

For the isochrone used in Figure \ref{n6791}, neglecting mass loss, the mass at the 
tip of the 
RGB is 1.12 $\Ms$. A 1 $\Ms$ He burning track is plotted to show the predicted level 
of the red clump in Figure \ref{n6791}. The He burning track provides a good fit to 
the V mag of the red clump but lies a few hundredths of a magnitude redder than the 
red clump in B--V for both color transformations. In V--I, the color of the red clump 
is in excellent agreement with the model.  In V--K, the model is bluer than the red 
clump by a few hundredths of a magnitude.

It should be noted that while the adopted reddening value is E(B--V) = 0.1, 
\citet{carney} found E(B--V) = 0.14 at $\feh$=+0.4 and the \citet{schlegel} dust maps 
yield E(B--V) = 0.155 for NGC~6791. These values are both higher than the adopted value 
of E(B--V) and would alter the isochrone fits shown in Figure \ref{n6791}.

If instead E(B--V) = 0.15 is adopted as the reddening then the isochrone fits require
either younger age or lower metallcity (or both).  Using the same isochrones (with 
$\feh_{init}$=+0.36) but with the larger reddening, a reasonably good fit is achieved
at 8 Gyr but with an increased distance modulus of $DM_V$ = 13.45.  Reducing $\feh$ to
+0.2 introduces a substantial mismatch between the observed CMD and the models along
the MS with the isochrones becoming bluer than the data at 2 mag below the MSTO.

Given that the adopted $\feh$ value provides a good fit to the entire CMD and agrees
with recent spectroscopic estimates by, for example, \citet{carr} and \citet{orig}, the 
isochrones show a clear preference for the lower reddening value of E(B--V)=0.1 used 
in Figure \ref{n6791}. Constraining the distance, reddening, composition, and age of 
NGC~6791 remains a thorny problem (see \citet{cha3} for a detailed discussion).  Because
of it high metallicity, a conclusive age determination of NGC~6791 will most likely 
require that the models follow the detailed abundance patterns found by spectroscopy
and not simply match the observed $\feh$.  For example, \citet{orig} found that $\afe \sim$ 0
but [C/Fe]=--0.35 in NGC~6791.

\subsubsection{Clusters in the SDSS ugriz system}

\begin{figure}
\plotone{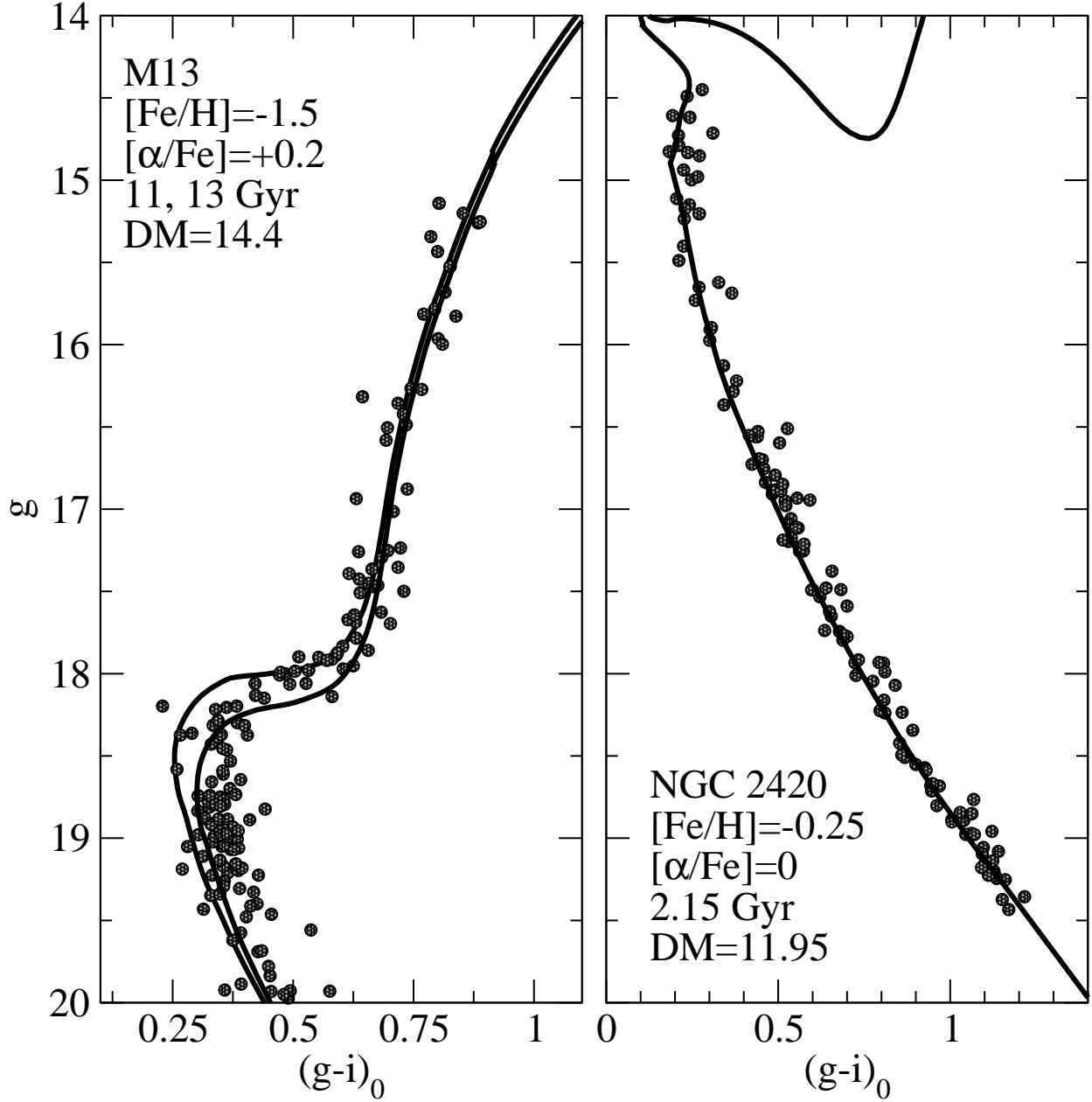}
\caption{The de-reddened g--i CMDs of M~13 and NGC~2420 from \citet{lee} plotted along with 
isochrones of appropriate age and metallicity transformed to the SDSS ugriz system.}
\label{sdss}
\end{figure}

\citet{lee} present spectroscopically selected cluster members for stars in a sample
of galactic globular and open clusters for purposes of validating the SEGUE Stellar 
Parameter Pipeline.  The selected stars from M~13 and NGC~2420 are used here because
they represent the cleanest available data sets of clusters on the SDSS ugriz system.
Though these sequences are quite narrow and well--defined, they are not as deep as 
allowed by SDSS, but are suitable for the purpose of demonstrating the performance 
of the isochrones.

Figure \ref{sdss} shows the g vs. (g--i)$_0$ CMDs of M~13 (left) and NGC~2420 (right) 
along with appropriate isochrones.  The photometry has been de-reddened by 
\citet{lee} using the \citet{schlegel} dust maps. Plotted on top of the M~13 data
in Figure \ref{sdss} are 
isochrones with $\feh$=-1.5, $\afe$=+0.2 at 11 and 13 Gyr  assuming a distance modulus 
of 14.4 in agreement with \citet{harris}.  The isochrones follow the locus
of stars on the MS and RGB well.  The only real discrepancy is the slope of the SGB  
but this is a common problem with synthetic color isochrones compared to metal poor 
systems.
 
Plotted on top of the NGC~2420 data in Figure \ref{sdss} is the same isochrone as in 
$\S$5.2.3 but now transformed to g vs. g--i plane. The agreement between the data and
 the isochrone is excellent on the MS but the observational constraints of the sample 
do not permit comparison with stars in the convective hook or past the MSTO.   

\subsection{Discussion}
Bluer synthetic colors, as represented by B--V in this section, continue to require 
adjustments in order to match observed CMDs as Figures \ref{m37} through \ref{n6791}
indicate.  The semi-empirical color transformation of VC03 are an improvement over the 
synthetic B--V color but seem unnecessary for V--I when used in conjuction with the 
DSEP isochrones.  While the synthetic colors and magnitudes are suitable analysis of 
observational data in bandpasses with central wavelengths longer than $\sim5000$ \AA, 
empirically adjusted colors ought to be used at shorter wavelengths.  In these redder
bands, the synthetic colors perform well throughout the CMD, including on the lower
main sequence where they perform better than the empirical colors.

The optical to near-infrared comparisons in V--J (M~37 only) and V--K of the open 
clusters are favorable for entire extent of the MS and the RGB in the older clusters 
but larger reddening than suggested by the adopted reddening curve is needed in some 
cases.

The location of He burning stars in the open clusters is matched by the models in the 
older clusters (M~67 and NGC~6791) but for the younger clusters the models predict the 
red clump to be fainter (NGC~2420) and/or redder (M~37 and the Hyades) than the data 
suggest. The general lack of evolved stars in the CMDs of both M~37 and the Hyades 
makes it difficult to speculate as to whether the models are intrinsically too cool 
or whether the colors are too red for a given $\Teff$.  

\section{Summary}

The Dartmouth Stellar Evolution Program has been used to compute a large grid of 
stellar evolution tracks.  A subset of these tracks and isochrones were presented by 
\citet{dotter} and are extended in this paper to include younger ages, higher 
metallicities, and transformations to multiple photometric systems. These models 
take advantage of the latest advances in radiative opacities, nuclear reactions rate, 
and the equation of state.

The Dartmouth Stellar Evolution Database comprises a set of stellar evolution 
tracks and isochrones transformed to several photometric systems along with a 
set of computer programs that allow for interpolation and construction of luminosity 
functions and synthetic horizontal branch models.  The database spans a wide range 
of compositions from $\feh$=--2.5 to +0.5, $\afe$=--0.2 to +0.8 (at $\feh \leq$0) or 
+0.2 (at $\feh >$0), and additional models with enhanced He for $\feh$=--2.5 to 0.  

The models were transformed to the observational plane using a synthetic color 
transformation based on \texttt{PHOENIX} model atmospheres for $\Teff <$10,000 K and 
\citet{cas} for hotter temperatures.  The synthetic color transformations are 
compared to the VC03 semi-empirical transformations in B--V and V--I. The synthetic 
colors perform well in V--I but are consistently too red in B--V.  As a general rule,
the synthetic colors should only be trusted in bandpasses equivalent to V or redder,
i.e. those with central wavelengths longer than $\sim5000$ \AA.  In these bands, the 
synthetic colors were found to perform well---and better than the empirical colors---
on the lower main sequence.
Synthetic photometry in the bluer bandpasses (those with central wavelengths shorter
than $\sim5000$ \AA) does not compare favorably with 
observations. Empirically adjusted magnitudes and colors should be favored for the 
bluer bandpasses whenever possible.

The isochrones were compared to isochrones from other groups at $\feh$=0 and +0.15 
as well as photometry from  open clusters and one globular cluster.  The isochrones 
extend to lower mass (0.1 $\Ms$) than those of other major isochrone libraries 
and provide reasonable fits to the MS of the clusters shown in $\S$5.

The Dartmouth Stellar Evolution Database should prove useful in studies of resolved 
stellar populations in the local Universe and is suitable for application to 
population synthesis and integrated light models.  The database is available through 
a dedicated website.

\acknowledgments
This work was supported by NSF grant AST-0094231.

The authors thank the anonymous referee for providing several helpful suggestions
that have improved the quality and utility of this paper.

AD wishes to thank Alan Irwin for his work on FreeEOS and for making it freely 
available, Arnold Boothroyd for sharing his subroutines to manage the OPAL Type 2 
opacity tables, Kevin Covey for helpful discussions regarding the SDSS photometric 
system, and Harvey Richer for advice regarding the HST photometric system.  
Thanks also to Aaron Grocholski, Ata Sarajedini, and Jason Kalirai for sharing their 
open cluster photometry.

The work of DJ and EB was supported in part by by NASA grants NAG5-3505 and NNG04GB36G, 
NSF grants AST-0307323 and AST-0707704, and US DOE Grant DE-FG02-07ER41517. This 
research used resources of the National Energy Research Scientific Computing Center 
(NERSC), which is supported by the Office of Science of the U.S. Department of Energy 
under Contract No. DE-AC03-76SF00098; and the H\"ochstleistungs Rechenzentrum Nord 
(HLRN).  We thank both these institutions for a generous allocation of computer time.

JF acknowledges support from NSF grant AST-0239590 and Grant No. EIA-0216178, Grant 
No. EPS-0236913 with matching support from the State of Kansas and the Wichita State 
University High Performance Computing Center.

\end{document}